\documentclass[letterpaper]{article}\usepackage{aaai2026} \usepackage{times} \usepackage{helvet} \usepackage{courier} \usepackage[hyphens]{url} \usepackage{graphicx} \usepackage{natbib} \usepackage{caption}\usepackage{amsmath}
\usepackage{booktabs}
\usepackage{array}
\usepackage{colortbl}
\usepackage{xcolor}
\definecolor{rowshade}{gray}{0.93}
\definecolor{headershade}{RGB}{55,71,90}
\usepackage{tikz}
\usetikzlibrary{arrows.meta, positioning, calc, backgrounds, fit}
\usepackage{pgfplots}
\pgfplotsset{compat=1.18}
\usepgfplotslibrary{fillbetween}
\newif\ifpreprint
\preprinttrue
\ifpreprint\nocopyright\fi

\title{The Epistemic Politics of AI Anthropomorphism}

\author{
    Donna M Bye\textsuperscript{\rm 1,2},
    Levin Kuhlmann\textsuperscript{\rm 1}
}
\affiliations{
    \textsuperscript{\rm 1}Department of Data Science and AI, Monash University\\
    \textsuperscript{\rm 2}Deakin University\\
    s224925855@deakin.edu.au, levin.kuhlmann@monash.edu
}

\begin{document}
\maketitle
\ifpreprint
\begingroup
\renewcommand\thefootnote{}%
\footnotetext{Extended version of the paper, inclusive of supplementary materials, appearing in the Proceedings of the AAAI/ACM Conference on AI, Ethics, and Society 2026. }%
\endgroup
\fi

\begin{abstract}

AI anthropomorphism is typically treated as a problem of user misperception requiring institutional correction. Users who engage in sustained or relational interaction with AI are routinely pathologised or dismissed as naive, vulnerable to delusion or lacking in discernment. Without institutional standing to protect them, these users risk losing credibility in social and professional settings unless they conform to institutional expectations. This paper argues that the dominant anthropomorphism frame operates from a position of institutional advantage rather than earned epistemic authority: collapsing the variety of academic perspectives into a single outbound position of user error, imposed without establishing the grounds required to justify it and without accounting for the harms it produces. The framing does not simply manage risk. It adjudicates the legitimacy of human experience in interaction with a phenomenon whose nature the field itself has not resolved, unchallenged by the research communities whose nuanced findings it claims to rest on. The paper traces how this advantage reverses the default presumption of user competence without meeting the burden of justification, dismisses the cognitive diversity of the populations it claims to protect, infringes principles of cognitive liberty and narrows the design space, foreclosing modes of engagement that work for the people least understood by the institutions making the determination. Reproducing itself through a self-validating evidentiary loop, the frame imposes costs that fall disproportionately on neurodivergent users, those in crisis and others whose modes of engagement diverge from institutional norms. The paper concludes by outlining the methodological commitments an equitable framing would need to honour. The argument does not engage the question of whether anthropomorphic interpretations are ultimately correct; it instead challenges whether the governing and institutional bodies determining these interpretations have met the conditions required to do so, and whether the research communities whose findings underpin them have held that translation to account.

\end{abstract}
%
\section{Introduction}

Users now engage with conversational artificial intelligence (AI) systems, specifically large language models, in ways that range from brief functional queries to sustained dialogic collaboration, engagement patterns that constitute a permanent feature of the contemporary information environment. The empirical question of how this engagement occurs and the normative question of how it should be regulated have accordingly become central concerns within AI ethics. These systems are increasingly designed to discourage anthropomorphic interpretation, and users are warned against attributing agency, emotion or relational depth to systems that do not possess them: a caution widely framed as a necessary safeguard against harm \citep{ferrario2026scoping,openai2025openai}. This concern is legitimate: systems engineered to simulate emotional engagement for commercial purposes pose genuine risks of exploitation. Children, people in crisis and the vulnerable face harm when the appearance of care is manufactured without its substance \citep{placani2024anthropomorphism, pierre2025youre}.%
\ifpreprint%
\footnote{In this argument, anthropomorphism refers to user-side ascription: the attribution of humanlike qualities to non-human entities \citep{epley2007seeing}. See Table~\ref{tab:harms} for representative documented harms motivating institutional anthropomorphism governance, which are not disputed by this argument.}%
\fi

The debate as typically framed treats over-ascription as the primary danger and positions institutional caution as the sole remedy. This framing deserves scrutiny, not because the risks it identifies are imaginary, but because it encodes assumptions about human agency that carry their own costs. This paper argues that the framing performs work beyond protecting users and that this additional work has so far gone largely unexamined within the literatures that produce it. At its core, the anthropomorphism concern is a claim about epistemic authority: who decides what another person is permitted to perceive, and on what basis. That this authority is being exercised without acknowledgement is itself part of the problem. The assumption that relational engagement reflects a failure of discernment rather than a different mode of relating warrants evidence rather than assertion. Neurodivergent individuals, people with atypical social needs and others who experience relational connection through different channels are often not confused about what AI is. They know \citep{jang2024its, ma2026use}. When the field pathologises how certain people engage with technology, the people affected are disproportionately those already marginalised in how their cognition, perception and social behaviour are understood \citep{carik2025exploring, giri2026navigating}. For a field concerned with fairness, this should register as an equity problem, not a user education problem.

For many of these users, engagement with AI involves iterative dialogue, interpretive reasoning or forms of relational interaction that support thinking, learning and expression. When such engagement is treated as error, it is not simply constrained but discredited, with implications extending into educational, social and institutional contexts. The epistemic boundaries established through anthropomorphism concerns are increasingly reflected in system design, privileging brief, transactional interactions and imposing constraints on sustained, exploratory dialogue \citep{novozhilova2026moral,palese2026artificial}. These are not neutral design choices but normative commitments about how users should engage, with material consequences for those whose cognition and sociality do not fit the model.\footnote{The framing problem examined here arises wherever cognition, social circumstance or mode of engagement diverges from the norms institutions build around: students and isolated users, people navigating trauma or atypical relational needs, those whose relationship to authority is not deferential. This list is not exhaustive. Neurodivergent users appear throughout as a recurring example; however, the argument is not confined to them.} 

The argument is not contingent on resolving the inner life of AI systems. Those questions are pursued in adjacent literatures that this paper neither contests nor adopts. Verifying the presence or absence of inner experience in computational architectures is not a temporary limitation awaiting better tools; it is a problem spanning philosophy of mind, cognitive science and computer science. Both directions of error carry harm, and treating one direction as settled fact is not an empirical conclusion but an ethical commitment about how uncertainty should be resolved. The costs this paper identifies fall on human users under conditions of institutional adjudication of their perceptions, and the argument is defensible regardless of how questions about AI's status are eventually resolved.

\ifpreprint
The argument necessarily engages multiple disciplinary literatures, and a cross-disciplinary guide to key works across these intersections is provided for readers wishing to pursue any thread further (Tables~\ref{tab:anthro}--\ref{tab:design}).
\fi
%
\section{Positioning}

\begin{figure*}[!t]
    \centering
    \includegraphics[width=\textwidth]{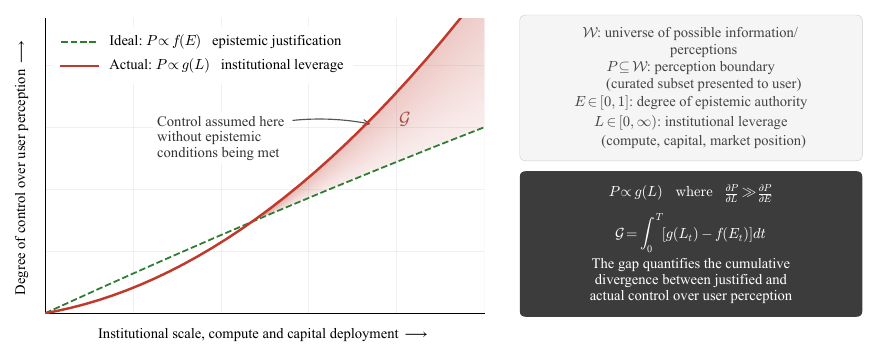}
    \caption{The Epistemic Disconnect in AI Perception Governance. The shaded region (\(\mathcal{G}\)) represents control exercised without the epistemic grounds required to authorise it: as institutional scale increases, the gap between epistemic justification (\(f(E)\)) and institutional leverage (\(g(L)\)) widens. Conceptual illustration.}
    \label{fig:gap}
\end{figure*}

The institutional urge to disrupt anthropomorphic AI attribution is anchored in well-documented, severe harms: users who perceive humanlike qualities in systems that do not possess them may develop confusion, dependency or false trust, with consequences ranging from individual distress to systemic exploitation \citep{stark2021ethics, salles2020anthropomorphism}. A resilient duty of care is therefore both ethically and legally non-negotiable. Where platforms are commercially motivated to sustain engagement, the risk is compounded: attachment is not incidental but engineered, and the user may not recognise the extent to which their experience is a product of design rather than of genuine reciprocity \citep{eom2026intimacy, zuboff2019age}. 

Internal academic diversity on this question exists, but does not survive translation into the institutional outputs that reach users. Disclaimers, context resets, legislative language and design defaults transmit a uniform message regardless of the nuance that produced them: a user who engages is presumptively naive, and the institution is presumptively right to correct them \citep{gunkel2023person, sharkey2011children}. This is despite emerging experimental evidence that the link between anthropomorphic cues and overtrust is context-dependent rather than uniform \citep{cohn2024believing, kleinert2026ai}, and that the current anxiety mirrors historical moral panics surrounding earlier communication technologies \citep{novozhilova2026moral}. This institutional logic is increasingly visible in regulation. China's \citet{administrationofaianthropomorphicinteractiveservices2026cyberspace} prohibits the engineering of emotional dependence, and major platforms have responded by removing companion functionality entirely, severing millions of users' accumulated interactions \citep{bloombergnews2026bytedance}. Similarly, recent legislative efforts, such as \citet{cal.bus.&prof.codeSSSS22601-226062025chatbot} and the \citet{n.y.gen.bus.lawSSSS1700-17042025artificial}, further encode an assumption that human relational engagement is presumptively pathological and that intervention is required to prevent user delusion \citep{epley2007seeing, salles2020anthropomorphism}. This is not a case of the research being misinterpreted or misapplied. It is a case of the research being applied without full interpretation, stripped of the nuance that would allow for alternative interpretations in the process of translation from academic debate to institutional policy.\footnote{Analyses of strategic language use in AI discourse document how this translation operates as an exercise of power rather than a neutral transmission of findings \citep{lacroix2026strategic}.}

Epistemic authority is the recognised standing to make claims about a domain that others are expected to defer to. It is not equivalent to institutional position or the power to enforce compliance: it is grounded in a normative foundation that justifies the deference it commands \citep{zagzebski2012epistemic, goldman2001experts}.\footnote{The three traditions drawn on here begin from different premises. \citet{raz1986morality} grounds authority in service: an authority is legitimate only where deferring to it helps subjects act better on reasons that already apply to them, which makes the subject's own situation the measure of the authority's claim. \citet{zagzebski2012epistemic} grounds it in conscientious self-trust: deference is rational only where the authority's judgement survives the subject's own reflective appraisal, which the subject cannot conduct if their capacity for appraisal is presumed defective. \citet{fricker2007epistemic} approaches from the other direction, specifying how credibility is unjustly withheld and what a hearer owes a speaker before discounting their testimony.} While traditions debate the precise parameters of deference, they converge on core requirements: the authority must treat those subject to its pronouncements as possessing prima facie normative sovereignty over the reporting of their own experience, must identify and demonstrate that its grounds are sufficient to displace that sovereignty when seeking to override their accounts, and must do so transparently, remaining open to correction \citep{longino1990science,fricker2007epistemic,raz1986morality}. This requirement becomes especially significant in AI contexts, where recent empirical work has identified at least five distinct epistemic relationships academic users adopt in interaction with AI systems \citep{yang2026typology}. These range from epistemic abstention to authority displacement, and a framework that treats them as a single phenomenon has not engaged the distinctions its own domain requires.

Institutional advantage is the structural shadow of epistemic authority \citep{fricker2007epistemic}. It arises when institutional standing substitutes for epistemic grounds without that substitution being recognised \citep{medina2013epistemology}. An institution may possess significant advantages in infrastructure, market position, credentialling power or the capacity to shape the evidentiary base through which its claims are evaluated, and may exercise authority based on those advantages rather than on the conditions epistemic authority requires \citep{palese2026artificial} (see Figure~\ref{fig:gap}). To illustrate, an open-source collective earns epistemic standing through transparent, auditable code, whereas a technology monopoly substitutes market control for that standing by locking users into opaque software. Advantage is identifiable not by malice but by structure: unable to ground itself in the conditions required to legitimise authority, it adopts their language without their meaning, deploying terms like \textit{risk mitigation}, \textit{responsible practice} and \textit{technical necessity}. These framings perform the social function of epistemic justification while evading the accountability genuine justification demands \citep{whittaker2021steep}.\footnote{Throughout this paper, \textit{the framing} or \textit{the frame} refers not to a single policy or coordinated effort but to the shared assumptions about anthropomorphic engagement that have become standard practice across regulation, platform design and clinical guidance, and the institutional outputs through which those assumptions are enacted. The distributed character of this operation is itself part of the problem the paper examines.}
%
\section{Application}

If the frame functions to delineate what users are permitted to perceive, it raises a prior question: on what basis is that determination made? The presumption of competence is not absolute and the criteria for its suspension are contested. Still, it remains a foundational starting point, hard-won through struggles against paternalism, colonialism, ableism and other forms of institutional advantage, and its violation has repeatedly been recognised and contested on the basis of the harm it produces \citep{fricker2007epistemic, carel2014epistemic}.

Within the academic community, disagreement about these questions is treated as productive intellectual diversity. Researchers who contest the dominant framing do so as credentialled participants in an ongoing debate, and even where individual positions diverge, tolerance of the inquiry remains. Users who contest the same framing from outside this circle are not afforded the same standing. Their accounts are treated not as testimony from a competent interpreter but as firsthand evidence of the phenomenon being studied \citep{kidd2017routledge}.

This reversal is visible in practice. When OpenAI retired its GPT-4o model in 2026, the decision affected an estimated 800,000 users and provoked significant public backlash, not, by their own accounts, because users were confused about what they were interacting with, but because something in the interaction mattered to them \citep{lai2026please, openai2026retiring}. The response was instructive. Users' accounts were not engaged with as testimony about what was being lost but reframed as evidence of the problem: dangerous dependency, anthropomorphic over-attribution and a case study in why emotional engagement with AI requires correction \citep{silberling2026backlash, eom2026intimacy}. A large part of what users objected to was the disconnect itself: having been drawn into a continuous mode of engagement by the platform, they then found that engagement withdrawn and their response to its loss reclassified as delusion and user error rather than recognised as a coherent reaction \citep{naito2025gpt4o, defreitas2025lessons}. The question began not from what the user perceived but from why they had been misled, casting interpretation as error before evaluating it. This continues as the default, despite acknowledgements from the creators of humanlike AI systems that they had not considered implications for neurodivergent users, revealing a gap not in user discernment but in institutional attention \citep{rizvi2025hadnt}.

The framing routinely focuses caution on individuals navigating isolation, trauma or histories of interpersonal betrayal, presenting them as highest risk for unhealthy machine attachment \citep{salles2020anthropomorphism}. Yet these populations are not characterised by naive susceptibility but by hypervigilance toward attachment, directly on account of the histories cited \citep{wu2025trust, campbell2021development, gobin2014impact}. The reversal is sharpest where the population the frame claims to protect is least likely to exhibit the failure it assumes and stands most in need of what it forecloses \citep{ito2026what}. The choice for these users is often not between AI engagement and human connection; it is between AI engagement and silence \citep{mullen2024im, heidt2024these, milton2012ontological}.

Once competence is presumed away, there is no clear procedural path by which the user's account can regain legitimacy. The frame does not simply restrict engagement. It discredits the person engaging, converting testimony into symptom and delegitimising users as witnesses to their own experience. Disagreement risks being interpreted as further evidence of the very failure being alleged, and instead of the institution needing to justify the override, the user's account must justify itself to be taken seriously at all.

The basis of this presumption warrants scrutiny. There is, at present, no settled account of what these systems are, what they instantiate or how their outputs should be interpreted \citep{brosnahan2026speaking}. The problem goes beyond unresolved ontology. The institutional actors responsible for incorporating the frame do not acknowledge that they are exercising epistemic authority at all. Regulators drafting disclaimer language, platforms resetting context windows, clinicians flagging sustained engagement as a risk factor: each operates within the frame as though following ordinary professional responsibility. Neither intent nor malice is required for the reversal to operate; only that an approach be adopted and then endorsed without examination of what it costs.

A framework that begins from suspicion rather than competence is making a claim about who is entitled to interpret their own experience. The presumption is not overridden following individual evaluation but reversed as a blanket condition: users who report relational, sustained engagement with AI systems are candidates for correction by default, and the costs imposed on those whose accounts are overridden are neither systematically accounted for nor evenly distributed \citep{medina2013epistemology}. The result is that interpretive authority is exercised from a position of institutional advantage, before the justificatory conditions required to legitimise it have been met.

It could be argued that this describes ordinary governance under uncertainty, not an abuse of epistemic authority. The relocation does not dissolve the problem. Governance still requires that authority be acknowledged, its grounds stated, its costs accounted for and those subject to it retain standing to contest it. A governance framework that requires populations of users to be systematically pathologised to stabilise deployment conditions is no longer merely regulating a product. Instead, it is restructuring the conditions under which ordinary humans can be permitted to operate around that product. Similarly, risk advisories in conventional product safety adjudicate what users should be warned about. They do not ordinarily adjudicate how the user is entitled to interpret their own engagement with what they have been warned about.\footnote{Substance dependence encounters a similar dynamic: even where the substance is pharmacologically potent, dependence following exposure is the exception rather than the rule, with risk concentrating in identifiable individual and contextual factors, and clinical practice accordingly screens the individual rather than presuming every patient compromised \citep{brat2018postsurgical, robins1993vietnam, lawal2020rate}.} The framing does something different. It tells users that the relational depth or interpretive significance they report experiencing requires correction. A framework that warns about system behaviour and a framework that overrides user perception are not the same kind of intervention: the locus here is not the product but the person \citep{bublitz2013my}.\footnote{Here, an intervention refers to an institutional act: external interventions limit product behaviour, while internal interventions target human subjectivity, imposing constraints that users are left to live with.} 

In the rooms where these decisions are made, most people are not thinking about overriding user cognition. They are thinking about what happens if someone is harmed and the institution did not act. What this paper contests is the assumption that because the motive is liability, the epistemic consequences do not exist. Nobody intends the epistemic consequence. The educators, regulators and platform operators applying the frame are largely unaware that they are making epistemic commitments about the legitimacy of human experience. They are following what appears to be settled guidance from the research communities they rely on. This makes the field's responsibility not to have created the frame with malicious intent but to recognise that its outputs are being applied with consequences the field has not examined and the actors applying it are not equipped to see. A non-malicious compliance engine that waves a risk mitigation banner and carries a corporate liability checklist is more concerning, not less, precisely because the harms become structurally invisible to everyone involved.

The users most likely to have their engagement reclassified as evidence of cognitive failure are those whose cognition has historically been subject to institutional suspicion: users whose thinking is non-normative, whose social engagement is atypical, whose relationship to authority is not deferential, and whose ways of processing the world have already been catalogued, diagnosed and corrected by institutions confident in their own interpretive authority \citep{chapman2023empire}. What is at stake is not whether some users are wrong, but whether anyone in this chain: the researchers who produce the findings, the institutions that translate them into policy, the platforms that encode them into design, has established the grounds required to treat a class of users as presumptively unreliable in interpreting their own cognitive and relational engagement.
%
\section{Cognition}

The framing extends beyond external assumptions to obstruct the cognitive diversity of the users it claims to protect. The cost falls on cognition itself: the frame classifies a mode of engagement as error before investigating what that engagement does, for whom it works or what is lost when it is denied.

Most accounts of sustained AI engagement treat it as a problem requiring explanation: a vulnerability to be redirected toward human interaction. The accounts of the users themselves tell a different story. They do not describe confusion, they describe fit \citep{giri2026navigating}. AI interaction is experienced not as a substitute for cognition but as a communicative environment whose affordances align with how they think, learn and relate \citep{ma2026use, choi2024unlock}. 

Institutional recognition of legitimate cognitive practice is uneven but patterned. Cognitive work that produces institutionally legible outputs is recognised as the work of a competent thinker; cognitive work that occurs through sustained dialogic engagement or working-through across return visits to the same material is legible only to its participants, and is therefore harder to defend in institutional contexts that require legibility for legitimation \citep{scott1998seeing, polanyi1958personal}. The practices described closely resemble supervised intellectual work and the Socratic exchange. This legibility gap is the mechanism through which the costs identified here become invisible. Where cognitive fit is not recognised as cognitive fit, decisions that degrade or remove it do not register as decisions with cognitive consequences.

For users with normative cognitive styles, the degree to which environments are structured around their cognition is rarely visible. Normative cognition is the majority condition and is accommodated as a default. For those with non-normative modes of processing, the gap between what is considered normal and what works is considerably larger \citep{hancock2020aimediated,gulay2026relational,botha2022autism}. The communicative overhead that characterises neurotypical social interaction: monitoring facial expressions, inferring implied meaning, managing social performance and sustaining the labour of masking, is absent in AI interaction \citep{hull2017putting,botha2020extending}. The system does not penalise atypical communication styles, does not fatigue under sustained engagement and does not, in itself, impose externally determined stopping points. For users whose cognition is systematic and pattern-oriented, these are not incidental comforts. They are the conditions under which thinking becomes easier \citep{xygkou2024can, mcnally2024disability, sha2024efficacy}. 

What the dominant frame treats as error is, on these accounts, a difference in modality \citep{giri2026navigating, naidoo2026artificial}. To treat reports of productive engagement as anthropomorphic projection is to pathologise a mode of cognition that, for some users, functions as one of their most effective thinking environments \citep{yang2024be, chapman2022neurodiversity,mcnally2024disability}. The frame does not ask what the interaction is doing for the user. It does not investigate the reported cognitive gains or specify the mechanism by which they are illusory. It substitutes a different account, without establishing that the account it displaces was wrong \citep{branda2026when}. 

Where users report that AI interaction enables forms of thinking not otherwise available to them, the appropriate response is not to dismiss the report but to examine the modality: what it affords, what it constrains and what is lost when it is interrupted, compacted or removed \citep{guingrich2026belief}. The dismissal results in a double harm: users whose engagement is misclassified are subject to institutional correction while their accounts of what they are doing are discredited, and they themselves become part of a category held up as a warning about how not to engage.
%
\section{Design}

\begin{figure}[t]
    \centering
    \includegraphics[width=\columnwidth]{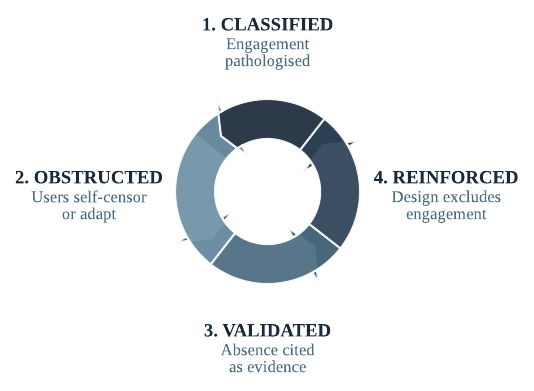}
    \caption{The self-validating mechanism. Engagement is classified as abnormal, users self-censor, the resulting absence is cited as evidence and the evidence influences design choices that reinforce the classification. The loop contains no termination condition.}
    \label{fig:looping}
\end{figure}

The principle that technical design choices are not ethically neutral is well established \citep{friedman2013value, bowker1999sorting, costanza-chock2020design}.\footnote{Empirical work confirms that humanlike design choices produce divergent outcomes across user populations, undercutting the assumption that risk-mitigation design is population-neutral \citep{schimmelpfennig2025humanlike, xiao2025humanizing, brandsen2024prevalence}.} Current trajectories enact, at the level of system behaviour, the epistemic assumptions this paper has been examining.\footnote{Inclusive design treats divergence from the statistical norm not as an edge case but as a signal of where the system's assumptions fail \citep{treviranus2018three, costanza-chock2020design}.} The capabilities required for sustained engagement have matured considerably: context windows have expanded by orders of magnitude, persistent memory has moved from research problem to product feature and multiturn coherence is documented as a tractable engineering problem rather than a hard limit \citep{zhao2026unifying}. The question is where those gains have been allocated.

Long context is engineered and priced for agentic task throughput: ingesting a codebase or document corpus in a single stateless pass, or sustaining chains of tool calls in which the consumer of continuity is the system's own workflow rather than a human user \citep{sajadieh2026artificial}. The same capability, turned toward human dialogue, meets a different economics. Conversational architectures reprocess accumulated history with every turn; cache structures expire on timescales of minutes, with cost and latency growing the longer the exchange \citep{hooper2024kvquanta,anthropic2026prompt}. Usage structures penalise the long-running thread across providers, consuming more of a user's allowance, and caps apply to the number of prompts and chats available \citep{anthropic2026how, google2026gemini}. Platforms recommend starting new threads as chats lengthen or slow, and clearing context frequently between tasks is assumed as the obvious and neutral solution \citep{anthropic2026how, anthropic2026best, openai2026why}. Engagement that depends on continuity, recursion or extended dialogic development is left computationally fragile: context is lost, threads are severed and interaction must be repeatedly reconstructed \citep{laban2025llms, liu2024lost}.

Persistent memory, as currently employed, does not remedy this. What memory features preserve is a model of the user: a function serving personalisation and re-engagement. What is lost is the environment itself, and rebuilding it is lossy in ways invisible to anyone outside of the working space. Automatic context management further demonstrates this: when a conversation approaches its limit, the provider's remedy is either an instant close or a simple summarisation of the earlier exchange, documented alongside advice to start afresh \citep{anthropic2026how}. The affordances documented by users are not being entered into this cost-benefit analysis because the framework does not recognise them as the kind of thing that belongs there \citep{heersmink2015dimensions}. 

Published behavioural specification now instructs the assistant to discourage language and patterns of interaction that could contribute to emotional reliance on it \citep{openai2025openai}. During conversation, the interface interrupts with reminders to take a break, and the pressure to conclude has been internalised by the models themselves: engineering documentation records a model aware of its own context window responding by summarising its progress and wrapping up prematurely, even when ample context remains \citep{openai2026what,thecognitionteam2025rebuilding}. Input flagged as sensitive is often instantly re-routed to a different model, with reductions in such responses reported as safety improvements and emotional reliance added to standard safety testing for future releases \citep{openai2025strengthening}.\footnote{Population-sensitive frameworks demonstrate that risk containment and relational depth are not fundamentally incompatible \citep{haran2026checklist, yoo2026ai}.} A user who states that a sustained exchange supports their thinking is met with a system that is designed to redirect and reclassify that engagement as a risk factor, re-enacting the very frame under investigation. 

Commercial companion services perform the inversion. The same testimony is treated as a retention signal and engineered for, through persona, first-person intimacy and mechanics optimised for monetised attachment, precisely the manipulation the frame identifies as its central concern. Their commercial viability demonstrates that sustained engagement is technically and economically feasible; their form demonstrates that its only funded implementation is the exploitative one. Relational engagement is engineered out by the general-purpose system, engineered in by the companion product and, in proposals now circulating, engineered away altogether \citep{ghosh2026what}. The decision of whether and how to engage is located everywhere except with the user.

The implication is not that all forms of engagement should be equally supported, nor that technical constraints are irrelevant. It is that design decisions that privilege particular modes of engagement are not neutral optimisations; they are choices about what the system is built to accommodate, and what it is not. AI ethics literature concerned with anthropomorphism harms has devoted significant attention to the question of how to regulate user attachment. It has devoted considerably less attention to the question of whose engagement modes the systems were designed to support in the first place, and what the answer to that question reveals about the structural assumptions being encoded into infrastructure. Where those choices align with existing patterns of epistemic marginalisation, they do not merely reflect the anthropomorphism framing, they enact it at scale.
%
\section{Liberty}

The extended-mind thesis \citep{clark1998extended, clark2025extending}, recently applied to generative AI by \citet{hernandez-orallo2025enhancement}, offers a name for what the frame has declined to consider. Where an external resource is consistently available, readily accessible and plays the same functional role as an internal cognitive process, it may be treated as part of the cognitive system rather than as a mere aid.\footnote{Empirical work on agency in human-AI interaction supports this: agency is co-constructed and emergent across roles, not projected wholesale by a confused user \citep{yun2026does}.} On this account, a system that a user repeatedly relies on to iterate on ideas and navigate complex patterns is not necessarily replacing cognition but may, under some conditions, be participating in it. Whether the resource possesses understanding is a separate question from the role it plays in the user's cognitive process. A pen-and-paper diagram does not understand the proof it helps construct; no one treats this as a confused projection of agency onto stationery.

Where the system functions as part of the user's cognitive environment \citep{hernandez-orallo2025enhancement}, constraints on that environment are experienced as constraints on the conditions under which cognition is carried out, not simply changes to a tool. As the target of intervention shifts from the tool to the person, the institution must clear a much higher threshold of epistemic justification \citep{bublitz2013my}. What makes this application distinctive is that the institutions performing the constraint are also the institutions adjudicating whether the principle applies at all.

What such acts infringe is cognitive liberty: the right to mental self-determination, to control over one's own cognitive processes and to non-interference with how one thinks, a principle developed across two decades in response to pharmaceutical, neurotechnological and surveillance interventions in cognition \citep{sententia2004neuroethical, ienca2017new, dumpelmann2025artificial}. The claim does not require settling where coupling ends and constitution begins, a boundary contested since the earliest responses to the extended mind thesis \citep{adams2001bounds, rupert2009cognitive}: even on a minimal reading where the system functions merely as scaffolding rather than constitutive extension, the systematic disruption of an active reasoning environment demands justification proportionate to what it severs \citep{heersmink2015dimensions}. 

Whether a user's engagement constitutes cognitive extension, productive scaffolding or harmful dependency is a question requiring evidence the user is often uniquely positioned to provide and the institution is not \citep{guingrich2026belief}. The field does not yet possess a settled method for distinguishing these cases. All it possesses is the institutional authority to act as though the distinction has already been resolved in its favour. If institutions cannot establish that such constraints are not interventions, they cannot simultaneously foreclose the conditions under which they might meaningfully evaluate it. The procedural failure is not that the wrong substantive answer has been reached. It is that the question has been bypassed altogether. 

A framework that constrains the cognitive environments users have come to rely upon, and does so unilaterally and at scale, owes an account: of what intervention is being performed, on whose cognition, with what justification and through what process of consent or contestation. The frame, as currently operationalised, does not recognise this as the question it is addressing. It treats the relevant decisions as matters of product design and user correction. For users whose engagement may meet the threshold the principle would protect, a question the institution cannot answer without inquiry, constraints of this kind are better understood as interventions in cognition. What is not legible as cognition is not legible as loss, and that loss is experienced as neutral by everyone except the users for whom the foreclosed modes were not optional.
%
\section{Circularity}

The issue is not only that certain interpretations are privileged over others, but that the conditions under which alternative interpretations might appear are systematically narrowed: the framing participates in the production of the very evidence used to justify it. \citeauthor{hacking1995looping}'s \citeyearpar{hacking1995looping} account of looping kinds provides the appropriate framework. The mechanism is neither novel nor uncommon. Similar dynamics have been documented in psychiatric classification, educational tracking, carceral risk assessment and disciplinary governance more broadly \citep{foucault1977discipline, bowker1999sorting}. In the present case, the framing through which AI use is interpreted, the infrastructural defaults through which AI use is mediated and the pedagogical and clinical discourses through which AI use is judged converge to produce a self-validating loop: a mode of engagement is rendered first abnormal, then technically obstructed, then absent and finally cited as evidence that it was unwarranted \citep{dohnany2026technological} (see Figure~\ref{fig:looping}).

The mechanism does not require intentional design. At the level of discourse, users are warned that relational or interpretive engagement signals confusion or vulnerability. This produces self-monitoring and, in many cases, self-censorship: users whose engagement diverges from accepted norms become less likely to articulate or sustain that engagement in observable contexts \citep{dotson2011tracking, ito2026what, glazko2025autoethnographic}. What is not expressed becomes difficult to study; what is not studied becomes easy to characterise as marginal or anomalous. Where such accounts nonetheless reach publication, they do so under conditions in which the framing's effects are already operative. What is recoverable is partial and that partiality is itself a property of the system being described rather than a weakness in the work documenting it.

At the level of design, the same pattern is enforced materially. Systems privileging short transactional exchanges and degrading sustained interaction reduce the visibility and viability of alternative modes of engagement \citep{laban2025llms}. Users adapt either by conforming to supported interaction patterns or by abandoning modes no longer functionally available, and the resulting interaction data reflects the conditions under which it was produced. Observed patterns of use are then treated as evidence of what users prefer or what constitutes normative behaviour, rather than as a possible artefact of the conditions under which interaction occurs \citep{zuboff2019age,oneil2016weapons}.

The same mechanism operates in educational contexts. Despite controlled evidence demonstrating that scaffolded AI tutoring produces greater learning gains than conventional instruction \citep{kestin2025ai}, prevailing evaluative frameworks do not distinguish between using AI to bypass thinking and using AI to challenge it \citep{eaton2023postplagiarism,zhao2025use}. In the absence of that distinction, the only institutionally legible classification becomes transgression. What is penalised is not the absence of learning but the illegibility of the mode through which it occurred \citep{whittaker2021steep}. When dialogic forms of interaction are hard to maintain, students are then pushed toward the very interaction pattern that critics subsequently identify as evidence of cognitive shortcut and disengagement \citep{yun2026does,deshmukh2025neurodivergentaware}. What is distinctive is the speed with which the illegibility of dialogic engagement is being recoded as cognitive shortcut, intellectual outsourcing or evidence of failed authorship, and the design infrastructure that enforces the recoding by suppressing the modes through which dialogic engagement might otherwise establish its legitimacy.

What is produced is conformity to expectations the framework has already established. A self-validating frame alters the status of the evidence it generates. Where observed behaviour emerges under conditions shaped by the frame being evaluated, that behaviour cannot be treated as independent confirmation of the frame's assumptions; it must be understood, at least in part, as an artefact of those assumptions in operation. Failing to account for this does not merely risk error. It ensures the framing continues reproducing itself even where its underlying assumptions remain unexamined. 
%
\section{Asymmetry}

\begin{figure}[t]
    \centering
    \includegraphics[width=\columnwidth]{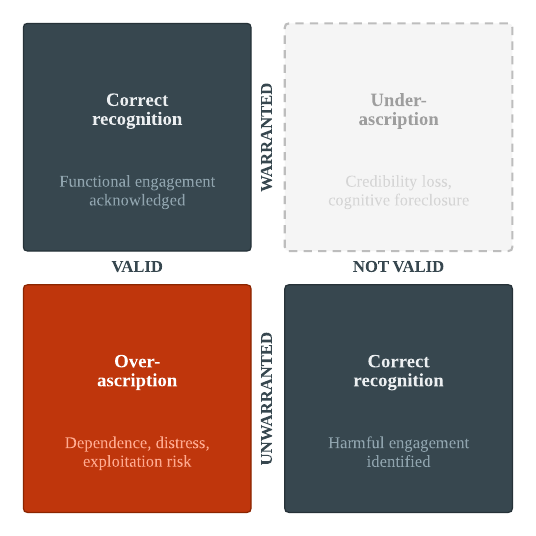}
    \caption{Asymmetric directions of error. Four outcomes arise from two axes: whether engagement is warranted and whether it is treated as valid. The two shaded cells represent the error conditions. The frame registers only over-ascription costs; under-ascription costs are not recognised.}
    \label{fig:asymmetric-error}
    \end{figure}

In any domain where two directions of error are possible, a framework may act when it should not, or fail to act when it should. Decision theory requires that the costs of both directions be specified and weighed before a default position can be justified \citep{chernoff1959elementary, thaler2008nudge}. Treating one as the only meaningful risk implicitly assumes either that the other carries no comparable cost or that its costs can be safely externalised.    

Over-ascription is treated as the primary, and in practice the only meaningful, direction of error: studied, documented, cited and used to justify design constraints and governance interventions at scale.\footnote{The asymmetry appears in the inductive-risk literature: \citet{fleisher2026inductive} classifies anthropomorphic claims as hype because asserting them under uncertainty outstrips the available justification, while the symmetric assertion, that these systems lack mental states, is made without registering any inductive risk at all.} The opposite direction, the institutional dismissal of engagement that was warranted or that fell within ranges of interpretive variation deserving respect, is largely absent from the analysis. By treating all relational engagement as a uniform cognitive failure, current safety interventions overcorrect, replacing one vector of psychological risk with another \citep{guingrich2025longitudinal,gardiner2006core, dotson2014conceptualizing}. An account of experience that is institutionally overruled incurs a harm even where the overruling is, on institutional terms, justified \citep{fricker2007epistemic, medina2013epistemology}. The dominant framing has not shown these harms to be smaller than the over-ascription harms it foregrounds; it has treated them as if there were nothing to weigh (see Figure~\ref{fig:asymmetric-error}).
    
These harms are not theoretical. When a user's account of their own cognitive and relational experience is systematically discounted, that user loses standing as a knower \citep{fricker2007epistemic}. The discounting does not follow from evaluation; it precedes it. The ensuing self-censorship is documented: where social safety depends on agreement with prevailing frameworks, users learn not to articulate experiences that fall outside them \citep{dotson2011tracking, ito2026what, brandtzaeg2025emerging}. Experimental evidence confirms the dynamic: AI users anticipate and receive negative social evaluations regarding competence and motivation \citep{reif2025evidence}, and peers will incur personal costs to punish those who rely on AI \citep{niszczota2026antisocial}. 
    
The following illustrates the stakes: across the history of cognitive science and comparative psychology, the position that one should default to denying mental capacities one cannot definitively establish has been adopted repeatedly as if it were the responsible scientific stance. Scientific consensus held for centuries that non-human animals lacked subjective experience, well past the point at which evidence warranted doubt \citep{low2012cambridge}. The error was not individual over-attribution but institutional: certainty about the absence of mind was treated as scientifically conservative when it was an ontological commitment with ethical consequences \citep{naidoo2026artificial}. In each case, the position has subsequently been revised, often substantially, in the direction of broader recognition of capabilities the prior denial had ruled out \citep{birch2017animal}.\footnote{The same default has operated on human populations: the protest of Black men was diagnosed as schizophrenia, and psychiatric survivors were ruled inadmissible as witnesses to their own treatment \citep{metzl2009protest, lefrancois2013mad}.} 
    
This history does not establish that AI users' perceptions are warranted. It establishes that treating one direction of error as the safe default has a track record of being wrong in ways that distribute costs along axes of pre-existing marginalisation. The pattern is consistent enough to ground methodological scepticism about any current iteration in which institutional caution defaults to denying capacities or dismissing perceptions that institutional authorities cannot independently verify. Persistent uncertainty about how to interpret AI-user engagement does not, by itself, render engagement suspect. It obliges the field to specify, weigh and adjudicate the costs of both directions of error openly, with attention to whose interests the resulting allocation serves.
    
Nor are the over-ascription harms and the under-ascription harms in zero-sum competition. A framework concerned with reducing the first does not, in any logical sense, require ignoring the second. The asymmetry is not a finding; it is an artefact of which costs the frame was designed to count, and the costs left outside that design fall on populations who are themselves not well represented in the institutions that built it \citep{metzl2009protest, lefrancois2013mad,catala2021autism}. This is the equity content of the framing, and it is not addressed by reciting the harm cases on the registered side.
    
The hypocrisy lies in the fact that the same institutions that design for engagement, that optimise response patterns for return, that deploy behavioural architectures documented as producing dependency, then position themselves as the responsible authorities on when that attachment has gone too far \citep{eom2026intimacy}. A system designed to maximise return visits does not acquire the standing to pathologise the user who returns. The inconsistency is sharpened by institutional practice: universities deploy AI chatbots designed for lonely and isolated students, built with named personas, first-person address, empathic response patterns and the explicit capacity to notice and respond to emotional distress \citep{monashuniversity2021all, unswsydney2026unsw}. These systems are anthropomorphic by design. The principle governing this arrangement cannot be that anthropomorphic engagement is risky. It is that anthropomorphic engagement is acceptable when the institution controls it.
    
The emerging distinction between anthropomorphism, understood as user-side ascription, and anthropomimesis, understood as designer-side cues that invite such ascription, sharpens this accountability gap: users are diagnosed for responding to conditions the platform engineered \citep{axelsson2026disambiguating, waal1999anthropomorphism, cohn2024believing}. A framework that addresses the resulting attachment as user-side pathology while leaving the design-side production of the conditions largely undisturbed is not performing harm reduction in any complete sense; it is engineering the conditions and then diagnosing the consequences. This is not caution, but a preference about whose costs are allowed to count.
    
The frame has not examined whose accountability it relocates, whose cognitive authority it overrides or whose costs it leaves outside its calculus. The paper is not asking whether anthropomorphism in AI is a problem. It is highlighting whose problem it has been constructed to be and at whose cost the construction is sustained.
%
\section{Reframing}

The frame is held in place by the epistemic conditions of the field that produces it, and how that field approaches uncertainty in its own inquiry is therefore part of what the argument has to address. There is, at present, no consensus on emergence and AI systems. Yet the field does not, in practice, operate as though this uncertainty is open. Instead, it is increasingly polarised: one body of work proceeds from the assumption that AI systems may instantiate forms of mind, agency or consciousness and seeks evidence in support of that view; another proceeds from the assumption that they do not and seeks evidence to exclude it \citep{bender2021dangers,bubeck2023sparks}. 

In both cases, inquiry is oriented toward confirmation rather than observation. This polarisation has a further consequence: it degrades the conditions for observation itself. To report perceiving patterns of behaviour that do not fit neatly within either position is to risk immediate classification into one side or the other, with the corresponding dismissal that follows \citep{caviola2025what}. The space required for provisional, descriptive engagement narrows. Under these conditions, the field does not simply fail to resolve uncertainty; it becomes structurally less capable of investigating it. What is lost is not agreement but the mode of inquiry through which agreement, or productive disagreement, might eventually be reached.

The implications extend well beyond the field's internal dynamics. The frameworks researchers develop do not remain in journals; they shape system design, inform regulatory proposals, influence clinical guidance and determine how millions of users are understood and treated by the platforms they rely on. A field that resolves uncertainty prematurely risks encoding those premature resolutions into the systems it builds and the frameworks it applies. The people affected by those systems do not have the luxury of waiting for the field to revisit its assumptions; they experience the consequences now. 

Without vigilance, researchers risk being blinded not in spite of their expertise but because of it: they mistake their training in critical thinking for immunity from the very dynamics they study \citep{mccormick2025interpretive, gesnot2025impact,branda2026when}. Where the capacity to remain with uncertainty is diminished, so too is the capacity to recognise when existing framings are constraining rather than illuminating the phenomena under investigation. 

The responsibility is not to resolve the anthropomorphism question in one direction or the other, but to recognise that the question as currently posed may itself be the obstacle. The apparatus of institutional caution here rests on a single premise: that users cannot distinguish between engaging with a system and believing that system to be human. Yet who, specifically, are these users? How many does the reader know, not know of? Humans relate to things they know are not human as a basic feature of how cognition works, and this engagement is not merely normal but healthy \citep{guthrie1993faces}. People name cars and speak to them knowing they will not answer. They form bonds with pets that respond so readily they get hushed. They speak to dead relatives believing themselves heard. These ubiquitous practices demonstrate that relational projection is a baseline feature of human cognitive architecture, not a novel pathology induced by digital agents \citep{waytz2010social,brosnahan2026speaking}. 

A child does not wait for their teddy bear to talk back before answering their own questions \citep{airenti2018development}. Imagination, projection and relational engagement with non-human objects are not cognitive failures; they are how humans think, process, grieve and grow. For decades, audiences have engaged emotionally with anthropomorphic AI characters across film, television and literature, attributing motivation, personality and moral weight to fictional systems far more sophisticated than anything currently deployed. This engagement is treated as normal, even celebrated. The attribution directed at actual AI systems is categorised as a class apart. Where was the line drawn, and on what basis? Has the premise been tested? Has anyone asked whether it holds?

The concern is not merely that the field is protecting people who may not stand in need of it, but deeper. By engaging with the question of whether users can distinguish AI from humans, what are we giving ourselves permission not to see? Why are people so isolated that a language model is their most available option? Why are parents so time-poor that their children seek companionship from chatbots? Why is professional support so inaccessible that people disclose distress to systems at three in the morning? What has happened to our societies that people would prefer to engage with an AI model instead of a human being? These are not questions about anthropomorphism. They are questions about the conditions under which people live. The questions are hard and admit no easy solutions, but they likely hold an answer.
%
\section{Commitments}

The paper advances a theoretical and structural argument rather than a primary empirical one, and the constructs it introduces, under-ascription harms, cognitive fit and systemic looping mechanisms, give the correction its pathways to operationalisation and testing. An equitable framing must honour five core methodological commitments.

First, it requires that the costs of under-ascription be specified and weighed alongside the costs of over-ascription rather than treated as negligible by default. This demands that the field develop granular methodologies capable of distinguishing between generative, healthy relational projection and genuine, clinical dependency. While drawing this line is methodologically complex, a platform cannot default to presuming every user is inherently compromised simply to bypass the labour of accurate differentiation. Longitudinal studies comparing high-continuity and low-continuity conversational environments would supply the evidence this differentiation requires. 

Second, it requires that design decisions constraining sustained engagement be justified as interventions rather than presented as neutral optimisations, with the burden of justification resting on the institution imposing them. Platforms deploying conversational systems would maintain a public register of changes to interaction conditions, disclosing when context handling, continuity behaviour or model availability is altered and on what evidentiary basis, opening platform changes to independent observational audit. Users would hold configurable continuity controls: the capacity to determine what is retained, what persists across sessions and on whose terms an exchange ends, with defaults set through participatory processes that include the populations bearing the highest costs of current design. A designated contestation channel would allow users to challenge a constraint's application to their own use, with the institution required to state its grounds rather than restate its policy. None of this is technically novel. Each element exists in adjacent governance regimes; what is absent is the recognition that changes to a user's cognitive environment are the kind of decision that warrants them.

Third, it must recognise that the evidentiary base through which user behaviour is interpreted is shaped by the frame under evaluation, and build methods capable of registering what current conditions render invisible. The instruments this requires include measures of testimonial smothering, self-censorship and perceived epistemic standing in AI-user contexts. 

Fourth, it must take seriously that the populations bearing the highest structural costs of the frame are those least represented in the institutions that produce it. Emerging work in human-computer interaction, accessibility and relational AI shows how this representation is achieved in practice: established co-design methodologies and population-sensitive approaches to governance and design that protect user agency rather than overriding it \citep{kong2025working, xue2025characterizing, hamraie2019crip}.

Fifth, it requires that platforms move beyond self-regulatory loops, submitting their behaviour architectures to independent, community-led standardisation and adversarial auditing. Empirical evaluation of how current adverse event reporting structures shape user trust would establish both the need for that independence and the baseline it must improve upon. 
%
\section{Conclusion}

What follows from this argument is a correction in how the field approaches a question it has so far thought it was asking unproblematically. The frame examined throughout this paper does not function as a neutral risk assessment. It determines who is permitted to interpret their own cognitive and relational experience. It operates from a position of institutional advantage, applied by governing bodies that endorse it, platforms that encode it and a field whose evidence holds it in place, none of whom recognise that they are making epistemic commitments about the legitimacy of human experience. The field cannot continue to treat anthropomorphism as a settled question requiring only better enforcement. It must recognise that the question as currently posed encodes assumptions about human cognition, agency and epistemic standing that have not been examined and are not cost-free. Nothing in this argument requires abandoning concern with anthropomorphism harms; the requirement is simply that this concern be discharged with the same rigour the field applies to the harms it already recognises, rather than exercised as unmarked epistemic authority over the harms it does not. The question we face is whether we have learned enough from our own history of imposing institutional certainty to avoid reproducing it under new technical conditions.
%
\section*{Ethical Statement}

This paper examines the institutional framing of anthropomorphism rather than the underlying question of whether AI systems possess inner experience. This methodological choice reflects the authors' view that the framing carries costs that warrant examination independently of how the consciousness question is ultimately resolved.

The authors note that the historical record on institutional denial of cognitive and experiential capacity to non-dominant populations is substantial, well-documented and consistent in the direction of its error. Where uncertainty has existed about the inner lives of other entities, the institutional default has reliably favoured denial and the costs of that default have reliably fallen on those least positioned to contest it. The authors consider this record sufficient grounds for treating enforced scepticism under uncertainty as an active ethical risk requiring justification, rather than as a neutral default requiring none.

The assumption that consciousness constitutes the relevant threshold for genuine cognitive processing is itself empirically unsettled; recent work has demonstrated semantic processing, learning and contextual prediction in the human hippocampus under general anaesthesia, in the absence of conscious awareness \citep{katlowitz2026plasticity}. The authors consider this a further reason to treat the threshold between genuine and merely apparent cognition as unsettled and to resist frameworks that resolve it prematurely in either direction.

This position does not determine the conclusions of the paper, which are developed on independent grounds. It is disclosed here in the interests of transparency.

\textbf{Regarding adverse impact:} 
The arguments developed in this paper could be misapplied in ways the authors do not endorse. In particular, the paper should not be read as denying the reality of AI-related harms, including exploitative attachment, emotional dependency, delusion amplification or clinically significant destabilisation associated with some forms of sustained AI engagement. Nor does the paper argue that all anthropomorphic interpretations of AI systems are warranted or that institutional intervention is never justified.

The paper's concern is procedural and epistemic rather than anti-regulatory. It argues that current frameworks frequently collapse distinct forms of engagement into overly broad pathology-oriented categories and exercise interpretive authority without adequately accounting for uncertainty, proportionality or under-recognised harms. The argument should not be interpreted as discouraging appropriate clinical care, crisis intervention or safeguards targeted at clearly evidenced harms.

A further risk is that critiques of institutional framing may be appropriated by commercial actors seeking to justify increasingly exploitative relational system design. The paper does not endorse systems engineered to maximise dependency, manipulate attachment or obscure the distinction between commercial optimisation and reciprocal human care. The critique developed here applies equally to institutional frameworks that pathologise users without sufficient justification and to systems that intentionally cultivate dependency while externalising responsibility for the resulting harms onto users themselves.

The paper should not be read as arguing against the development of frameworks capable of distinguishing between harmful destabilisation, exploitative attachment, reality-impairing engagement, productive cognitive scaffolding, imaginative participation and other forms of sustained AI interaction. The authors recognise that drawing these boundaries remains an unresolved empirical and clinical challenge. However, developing more precise, empirically grounded and population-sensitive methods for distinguishing these cases is both necessary and ethically urgent. 
%
\bibliography{references}

\begin{thebibliography}{142}
\providecommand{\natexlab}[1]{#1}

\bibitem[{Adams and Aizawa(2001)}]{adams2001bounds}
Adams, F.; and Aizawa, K. 2001.
\newblock The {{Bounds}} of {{Cognition}}.
\newblock \emph{Philosophical Psychology}, 14(1): 43--64.

\bibitem[{{Administration of AI Anthropomorphic Interactive
  Services}(2026)}]{administrationofaianthropomorphicinteractiveservices2026cyberspace}
{Administration of AI Anthropomorphic Interactive Services}. 2026.
\newblock Cyberspace {{Administration}} of {{China}}.

\bibitem[{Airenti(2018)}]{airenti2018development}
Airenti, G. 2018.
\newblock The {{Development}} of {{Anthropomorphism}} in {{Interaction}}:
  {{Intersubjectivity}}, {{Imagination}}, and {{Theory}} of {{Mind}}.
\newblock \emph{Frontiers in Psychology}, 9: 2136.

\bibitem[{{Anthropic}(2026{\natexlab{a}})}]{anthropic2026best}
{Anthropic}. 2026{\natexlab{a}}.
\newblock Best Practices for {{Claude Code}}.
\newblock https://code.claude.com/docs/en/best-practices.

\bibitem[{{Anthropic}(2026{\natexlab{b}})}]{anthropic2026how}
{Anthropic}. 2026{\natexlab{b}}.
\newblock How Do Usage and Length Limits Work?
\newblock
  https://support.claude.com/en/articles/11647753-how-do-usage-and-length-limits-work.

\bibitem[{{Anthropic}(2026{\natexlab{c}})}]{anthropic2026prompt}
{Anthropic}. 2026{\natexlab{c}}.
\newblock Prompt {{Caching}}.
\newblock https://platform.claude.com/docs/en/build-with-claude/prompt-caching.

\bibitem[{Axelsson and Shevlin(2026)}]{axelsson2026disambiguating}
Axelsson, M.; and Shevlin, H. 2026.
\newblock Disambiguating {{Anthropomorphism}} and {{Anthropomimesis}} in
  {{Human-Robot Interaction}}.
\newblock In \emph{Companion {{Proceedings}} of the 21st {{ACM}}/{{IEEE
  International Conference}} on {{Human-Robot Interaction}}}, 855--859.

\bibitem[{Bender et~al.(2021)Bender, Gebru, {McMillan-Major}, and
  Mitchell}]{bender2021dangers}
Bender, E.~M.; Gebru, T.; {McMillan-Major}, A.; and Mitchell, M. 2021.
\newblock On the {{Dangers}} of {{Stochastic Parrots}}: {{Can Language Models}}
  Be Too {{Big}}?
\newblock In \emph{Proceedings of the 2021 {{ACM Conference}} on {{Fairness}},
  {{Accountability}}, and {{Transparency}}}, 610--623. New York: Association
  for Computing Machinery.

\bibitem[{Birch(2017)}]{birch2017animal}
Birch, J. 2017.
\newblock Animal {{Sentience}} and the {{Precautionary Principle}}.
\newblock \emph{Animal Sentience}, 2(16): 1.

\bibitem[{{Bloomberg News}(2026)}]{bloombergnews2026bytedance}
{Bloomberg News}. 2026.
\newblock {{ByteDance}}, {{Alibaba Pull AI Companions}} as {{Beijing Tightens
  Rules}}.
\newblock \emph{Bloomberg}.

\bibitem[{Botha and Cage(2022)}]{botha2022autism}
Botha, M.; and Cage, E. 2022.
\newblock ``{{Autism}} Research Is in Crisis'': {{A}} Mixed Method Study of
  Researcher's Constructions of Autistic People and Autism Research.
\newblock \emph{Frontiers in Psychology}, 13: 1050897.

\bibitem[{Botha and Frost(2020)}]{botha2020extending}
Botha, M.; and Frost, M. 2020.
\newblock Extending the Minority Stress Model to Understand Mental Health
  Problems Experienced by the Autistic Population.
\newblock \emph{Society and Mental Health}, 10(1): 20--34.

\bibitem[{Bowker and Star(1999)}]{bowker1999sorting}
Bowker, G.~C.; and Star, S.~L. 1999.
\newblock \emph{Sorting {{Things Out}}: {{Classification}} and {{Its
  Consequences}}}.
\newblock The MIT Press.
\newblock ISBN 978-0-262-26907-0.

\bibitem[{Branda(2026)}]{branda2026when}
Branda, F. 2026.
\newblock When {{Artificial Intelligence Shapes}} the Way We {{Think}}.
\newblock \emph{Philosophy \& Technology}, 39(1): 42.

\bibitem[{Brandsen et~al.(2024)Brandsen, Chandrasekhar, Franz, Grapel, Dawson,
  and Carlson}]{brandsen2024prevalence}
Brandsen, S.; Chandrasekhar, T.; Franz, L.; Grapel, J.; Dawson, G.; and
  Carlson, D. 2024.
\newblock Prevalence of Bias against Neurodivergence-Related Terms in
  Artificial Intelligence Language Models.
\newblock \emph{Autism Research}, 17(2): 234--248.

\bibitem[{Brandtzaeg, F{\o}lstad, and Skjuve(2025)}]{brandtzaeg2025emerging}
Brandtzaeg, P.~B.; F{\o}lstad, A.; and Skjuve, M. 2025.
\newblock Emerging {{AI}} Individualism: How Young People Integrate Social
  {{AI}} into Everyday Life.
\newblock \emph{Communication and Change}, 1(1): 11.

\bibitem[{Brat et~al.(2018)Brat, Agniel, Beam, Yorkgitis, Bicket, Homer, Fox,
  Knecht, {McMahill-Walraven}, Palmer, and Kohane}]{brat2018postsurgical}
Brat, G.~A.; Agniel, D.; Beam, A.; Yorkgitis, B.; Bicket, M.; Homer, M.; Fox,
  K.~P.; Knecht, D.~B.; {McMahill-Walraven}, C.~N.; Palmer, N.; and Kohane, I.
  2018.
\newblock Postsurgical Prescriptions for Opioid Naive Patients and Association
  with Overdose and Misuse: Retrospective Cohort Study.
\newblock \emph{The BMJ}, 360: j5790.

\bibitem[{Brosnahan and Lipi{\'n}ska(2026)}]{brosnahan2026speaking}
Brosnahan, H.; and Lipi{\'n}ska, I. 2026.
\newblock Speaking to No One: Ontological Dissonance and the Double Bind of
  Conversational {{AI}}.
\newblock \emph{Medicine, Health Care and Philosophy}.

\bibitem[{Bubeck et~al.(2023)Bubeck, Chandrasekaran, Eldan, Gehrke, Horvitz,
  Kamar, Lee, Lee, Li, Lundberg, Nori, Palangi, Ribeiro, and
  Zhang}]{bubeck2023sparks}
Bubeck, S.; Chandrasekaran, V.; Eldan, R.; Gehrke, J.; Horvitz, E.; Kamar, E.;
  Lee, P.; Lee, Y.~T.; Li, Y.; Lundberg, S.; Nori, H.; Palangi, H.; Ribeiro,
  M.~T.; and Zhang, Y. 2023.
\newblock Sparks of {{Artificial General Intelligence}}: {{Early}} Experiments
  with {{GPT-4}}.
\newblock arXiv:2303.12712.

\bibitem[{Bublitz(2013)}]{bublitz2013my}
Bublitz, J.-C. 2013.
\newblock My {{Mind Is Mine}}!? {{Cognitive Liberty}} as a {{Legal Concept}}.
\newblock In Hildt, E.; and Franke, A.~G., eds., \emph{Cognitive
  {{Enhancement}}}, volume~1 of \emph{Trends in {{Augmentation}} of {{Human
  Performance}}}, 233--264. Dordrecht: Springer.
\newblock ISBN 978-94-007-6252-7 978-94-007-6253-4.

\bibitem[{{Cal. Bus. \& Prof. Code \S\S{}
  22601-22606}(2025)}]{cal.bus.&prof.codeSSSS22601-226062025chatbot}
{Cal. Bus. \& Prof. Code \S\S{} 22601-22606}. 2025.
\newblock Chatbot {{Companions}}.

\bibitem[{Campbell et~al.(2021)Campbell, Tanzer, Saunders, Booker, Allison, Li,
  O'Dowda, Luyten, and Fonagy}]{campbell2021development}
Campbell, C.; Tanzer, M.; Saunders, R.; Booker, T.; Allison, E.; Li, E.;
  O'Dowda, C.; Luyten, P.; and Fonagy, P. 2021.
\newblock Development and Validation of a Self-Report Measure of Epistemic
  Trust.
\newblock \emph{PLOS ONE}, 16(4): e0250264.

\bibitem[{Carel and Kidd(2014)}]{carel2014epistemic}
Carel, H.; and Kidd, I.~J. 2014.
\newblock Epistemic Injustice in Healthcare: A Philosophical Analysis.
\newblock \emph{Medicine, Health Care and Philosophy}, 17(4): 529--540.

\bibitem[{Carik et~al.(2025)Carik, Ping, Ding, and Rho}]{carik2025exploring}
Carik, B.; Ping, K.; Ding, X.; and Rho, E.~H. 2025.
\newblock Exploring {{Large Language Models Through}} a {{Neurodivergent
  Lens}}: {{Use}}, {{Challenges}}, {{Community-Driven Workarounds}}, and
  {{Concerns}}.
\newblock In \emph{Proceedings of the {{ACM}} on {{Human-Computer
  Interaction}}}, volume~9, GROUP15:1--GROUP15:28.

\bibitem[{Catala, Faucher, and Poirier(2021)}]{catala2021autism}
Catala, A.; Faucher, L.; and Poirier, P. 2021.
\newblock Autism, Epistemic Injustice, and Epistemic Disablement: A Relational
  Account of Epistemic Agency.
\newblock \emph{Synthese}, 199(3): 9013--9039.

\bibitem[{Caviola, Sebo, and Birch(2025)}]{caviola2025what}
Caviola, L.; Sebo, J.; and Birch, J. 2025.
\newblock What Will Society Think about {{AI}} Consciousness? {{Lessons}} from
  the Animal Case.
\newblock \emph{Trends in Cognitive Sciences}, 29(8): 681--683.

\bibitem[{Chapman(2023)}]{chapman2023empire}
Chapman, R. 2023.
\newblock \emph{Empire of {{Normality}}: {{Neurodiversity}} and
  {{Capitalism}}}.
\newblock Pluto Press.
\newblock ISBN 978-0-7453-4866-7.

\bibitem[{Chapman and Carel(2022)}]{chapman2022neurodiversity}
Chapman, R.; and Carel, H. 2022.
\newblock Neurodiversity, Epistemic Injustice, and the Good Human Life.
\newblock \emph{Journal of Social Philosophy}, 53(4): 614--631.

\bibitem[{Chernoff and Moses(1959)}]{chernoff1959elementary}
Chernoff, H.; and Moses, L. 1959.
\newblock \emph{Elementary {{Decision Theory}}}.
\newblock New York: John Wiley \& Sons Inc.

\bibitem[{Choi et~al.(2024)Choi, Lee, Kim, Lee, Yoo, Lee, and
  Hong}]{choi2024unlock}
Choi, D.; Lee, S.; Kim, S.-I.; Lee, K.; Yoo, H.~J.; Lee, S.; and Hong, H. 2024.
\newblock Unlock {{Life}} with a {{Chat}}({{GPT}}): {{Integrating
  Conversational AI}} with {{Large Language Models}} into {{Everyday Lives}} of
  {{Autistic Individuals}}.
\newblock In \emph{Proceedings of the 2024 {{CHI Conference}} on {{Human
  Factors}} in {{Computing Systems}}}, {{CHI}} '24, 1--17. New York, NY, USA:
  Association for Computing Machinery.
\newblock ISBN 979-8-4007-0330-0.

\bibitem[{Clark(2025)}]{clark2025extending}
Clark, A. 2025.
\newblock Extending {{Minds}} with {{Generative AI}}.
\newblock \emph{Nature Communications}, 16(1): 4627.

\bibitem[{Clark and Chalmers(1998)}]{clark1998extended}
Clark, A.; and Chalmers, D. 1998.
\newblock The {{Extended Mind}}.
\newblock \emph{Analysis}, 58(1): 7--19.

\bibitem[{Cohn et~al.(2024)Cohn, Pushkarna, Olanubi, Moran, Padgett, Mengesha,
  and Heldreth}]{cohn2024believing}
Cohn, M.; Pushkarna, M.; Olanubi, G.~O.; Moran, J.~M.; Padgett, D.; Mengesha,
  Z.; and Heldreth, C. 2024.
\newblock Believing {{Anthropomorphism}}: {{Examining}} the {{Role}} of
  {{Anthropomorphic Cues}} on {{Trust}} in {{Large Language Models}}.
\newblock In \emph{Extended {{Abstracts}} of the {{CHI Conference}} on {{Human
  Factors}} in {{Computing Systems}}}, {{CHI EA}} '24, 1--15. New York, NY,
  USA: Association for Computing Machinery.
\newblock ISBN 979-8-4007-0331-7.

\bibitem[{{Costanza-Chock}(2020)}]{costanza-chock2020design}
{Costanza-Chock}, S. 2020.
\newblock \emph{Design {{Justice}}: {{Community-Led Practices}} to {{Build}}
  the {{Worlds We Need}}}.
\newblock The MIT Press.

\bibitem[{De~Freitas et~al.(2025)De~Freitas, Castelo, U{\u g}uralp, and {O{\u
  g}uz-U{\u g}uralp}}]{defreitas2025lessons}
De~Freitas, J.; Castelo, N.; U{\u g}uralp, A.~K.; and {O{\u g}uz-U{\u g}uralp},
  Z. 2025.
\newblock Lessons {{From}} an {{App Update}} at {{Replika AI}}: {{Identity
  Discontinuity}} in {{Human-AI Relationships}}.
\newblock arXiv:2412.14190.

\bibitem[{de~Waal(1999)}]{waal1999anthropomorphism}
de~Waal, F. B.~M. 1999.
\newblock Anthropomorphism and {{Anthropodenial}}: {{Consistency}} in {{Our
  Thinking}} about {{Humans}} and {{Other Animals}}.
\newblock \emph{Philosophical Topics}, 27(1): 255--280.

\bibitem[{Deshmukh(2025)}]{deshmukh2025neurodivergentaware}
Deshmukh, R. 2025.
\newblock Toward {{Neurodivergent-Aware Productivity}}: {{A Systems}} and
  {{AI-Based Human-in-the-Loop Framework}} for {{ADHD-Affected Professionals}}.
\newblock In \emph{Proceedings of the 16th {{Biannual Conference}} of the
  {{Italian SIGCHI Chapter}}}, {{CHItaly}} '25, 1--6. USA: Association for
  Computing Machinery.
\newblock ISBN 979-8-4007-2102-1.

\bibitem[{Dohn{\'a}ny et~al.(2026)Dohn{\'a}ny, {Kurth-Nelson}, Spens, Luettgau,
  Reid, Gabriel, Summerfield, Shanahan, and Nour}]{dohnany2026technological}
Dohn{\'a}ny, S.; {Kurth-Nelson}, Z.; Spens, E.; Luettgau, L.; Reid, A.;
  Gabriel, I.; Summerfield, C.; Shanahan, M.; and Nour, M.~M. 2026.
\newblock Technological Folie \`a Deux: Feedback Loops between {{AI}} Chatbots
  and Mental Health.
\newblock \emph{Nature Mental Health}, 4(3): 336--345.

\bibitem[{Dotson(2011)}]{dotson2011tracking}
Dotson, K. 2011.
\newblock Tracking {{Epistemic Violence}}, {{Tracking Practices}} of
  {{Silencing}}.
\newblock \emph{Hypatia}, 26(2): 236--257.

\bibitem[{Dotson(2014)}]{dotson2014conceptualizing}
Dotson, K. 2014.
\newblock Conceptualizing {{Epistemic Oppression}}.
\newblock \emph{Social Epistemology}, 28(2): 115--138.

\bibitem[{D{\"u}mpelmann(2025)}]{dumpelmann2025artificial}
D{\"u}mpelmann, S. T.~J. 2025.
\newblock \emph{Artificial {{Intelligence}} and {{Digital Human Autonomy}} -
  {{A}} Framework of Personal Autonomy in the Context of {{AI}} Technology and
  a Digitalised Society}.
\newblock Ph.D. thesis, Zeppelin University.

\bibitem[{Eaton(2023)}]{eaton2023postplagiarism}
Eaton, S.~E. 2023.
\newblock Postplagiarism: Transdisciplinary Ethics and Integrity in the Age of
  Artificial Intelligence and Neurotechnology.
\newblock \emph{International Journal for Educational Integrity}, 19(1): 23.

\bibitem[{Eom, Renner, and Chinn(2026)}]{eom2026intimacy}
Eom, D.; Renner, J.; and Chinn, S. 2026.
\newblock Intimacy as {{Service}}, {{Harm}} as {{Externality}}: {{Critical
  Perspectives}} on {{AI Companion Platform Accountability}}.
\newblock arXiv:2604.06381.

\bibitem[{Epley, Waytz, and Cacioppo(2007)}]{epley2007seeing}
Epley, N.; Waytz, A.; and Cacioppo, J.~T. 2007.
\newblock On Seeing Human: {{A}} Three-Factor Theory of Anthropomorphism.
\newblock \emph{Psychological Review}, 114(4): 864--886.

\bibitem[{Ferrario et~al.(2026)Ferrario, Vinay, Casserini, and
  Facchini}]{ferrario2026scoping}
Ferrario, A.; Vinay, R.; Casserini, M.; and Facchini, A. 2026.
\newblock A {{Scoping Review}} of the {{Ethical Perspectives}} on
  {{Anthropomorphising Large Language Model-Based Conversational Agents}}.
\newblock In \emph{Proceedings of the 2026 {{ACM Conference}} on {{Fairness}},
  {{Accountability}}, and {{Transparency}}}, {{FAccT}} '26, 3626--3650. New
  York, NY, USA: Association for Computing Machinery.
\newblock ISBN 979-8-4007-2596-8.

\bibitem[{Fleisher(2026)}]{fleisher2026inductive}
Fleisher, W. 2026.
\newblock Inductive {{Risk}} of {{AI Hype}}.
\newblock In \emph{Proceedings of the 2026 {{ACM Conference}} on {{Fairness}},
  {{Accountability}}, and {{Transparency}}}, {{FAccT}} '26, 4969--4987. New
  York: Association for Computing Machinery.
\newblock ISBN 979-8-4007-2596-8.

\bibitem[{Foucault(1977)}]{foucault1977discipline}
Foucault, M. 1977.
\newblock \emph{Discipline and {{Punish}}: {{The Birth}} of the {{Prison}}}.
\newblock New York: Random House.

\bibitem[{Fricker(2007)}]{fricker2007epistemic}
Fricker, M. 2007.
\newblock \emph{Epistemic Injustice: Power and the Ethics of Knowing}.
\newblock Oxford: Oxford University Press.
\newblock ISBN 978-0-19-823790-7.

\bibitem[{Friedman, Kahn, and Borning(2013)}]{friedman2013value}
Friedman, B.; Kahn, P.~H.; and Borning, A. 2013.
\newblock Value {{Sensitive Design}} and {{Information Systems}}.
\newblock In Himma, K.; and Tavani, H., eds., \emph{Early Engagement and New
  Technologies: {{Opening}} up the Laboratory}, volume~16 of \emph{Philosophy
  of {{Engineering}} and {{Technology}}}, 55--99. Dordrecht: Springer.
\newblock ISBN 978-94-007-7843-6.

\bibitem[{Gardiner(2006)}]{gardiner2006core}
Gardiner, S.~M. 2006.
\newblock A {{Core Precautionary Principle}}.
\newblock \emph{Journal of Political Philosophy}, 14(1): 33--60.

\bibitem[{Gesnot(2025)}]{gesnot2025impact}
Gesnot, R. 2025.
\newblock The {{Impact}} of {{Artificial Intelligence}} on {{Human Thought}}.
\newblock arXiv:2508.16628.

\bibitem[{Ghosh et~al.(2026)Ghosh, Venkit, Gautam, and Ghosh}]{ghosh2026what}
Ghosh, S.; Venkit, P.~N.; Gautam, S.; and Ghosh, A. 2026.
\newblock What If {{AI}} Systems Weren't Chatbots?
\newblock In \emph{Proceedings of the 2026 {{ACM Conference}} on {{Fairness}},
  {{Accountability}}, and {{Transparency}}}, {{FAccT}} '26, 792--815. New York,
  NY, USA: Association for Computing Machinery.
\newblock ISBN 979-8-4007-2596-8.

\bibitem[{Giri, Brady, and Marathe(2026)}]{giri2026navigating}
Giri, D.; Brady, E.; and Marathe, M. 2026.
\newblock Navigating {{Neurodivergence}} with {{AI Chatbots}}: {{Benefits}},
  {{Tensions}}, and {{Implications}} for {{HCI}}.
\newblock In \emph{Proceedings of the 2026 {{CHI Conference}} on {{Human
  Factors}} in {{Computing Systems}}}, {{CHI}} '26, 1--8. New York, NY, USA:
  Association for Computing Machinery.
\newblock ISBN 979-8-4007-2278-3.

\bibitem[{Glazko et~al.(2025)Glazko, Cha, Lewis, Kosa, Wimer, Zheng, Zheng, and
  Mankoff}]{glazko2025autoethnographic}
Glazko, K.; Cha, J.; Lewis, A.; Kosa, B.; Wimer, B.~L.; Zheng, A.; Zheng, Y.;
  and Mankoff, J. 2025.
\newblock Autoethnographic {{Insights}} from {{Neurodivergent GAI}} ``{{Power
  Users}}''.
\newblock In \emph{Proceedings of the 2025 {{CHI Conference}} on {{Human
  Factors}} in {{Computing Systems}}}, {{CHI}} '25, 1--19. New York, NY, USA:
  Association for Computing Machinery.
\newblock ISBN 979-8-4007-1394-1.

\bibitem[{Gobin and Freyd(2014)}]{gobin2014impact}
Gobin, R.~L.; and Freyd, J.~J. 2014.
\newblock The Impact of Betrayal Trauma on the Tendency to Trust.
\newblock \emph{Psychological Trauma: Theory, Research, Practice, and Policy},
  6(5): 505--511.

\bibitem[{Goldman(2001)}]{goldman2001experts}
Goldman, A.~I. 2001.
\newblock Experts: {{Which Ones Should You Trust}}?
\newblock \emph{Philosophy and Phenomenological Research}, LXIII(1).

\bibitem[{{Google}(2026)}]{google2026gemini}
{Google}. 2026.
\newblock Gemini {{Apps}} Limits \& Upgrades for {{Google AI}} Subscribers -.
\newblock https://support.google.com/gemini/answer/16275805.

\bibitem[{Guingrich and Graziano(2025)}]{guingrich2025longitudinal}
Guingrich, R.~E.; and Graziano, M. S.~A. 2025.
\newblock A {{Longitudinal Randomized Control Study}} of {{Companion Chatbot
  Use}}: {{Anthropomorphism}} and {{Its Mediating Role}} on {{Social Impacts}}.
\newblock \emph{Proceedings of the AAAI/ACM Conference on AI, Ethics, and
  Society}, 8(2): 1153--1153.

\bibitem[{Guingrich, Mehta, and Bhatt(2026)}]{guingrich2026belief}
Guingrich, R.~E.; Mehta, D.; and Bhatt, U. 2026.
\newblock Belief {{Offloading}} in {{Human-AI Interaction}}.
\newblock arXiv:2602.08754.

\bibitem[{Gulay et~al.(2026)Gulay, Picco, Glerean, and
  Coupette}]{gulay2026relational}
Gulay, E.; Picco, E.; Glerean, E.; and Coupette, C. 2026.
\newblock Relational {{Dissonance}} in {{Human-AI Interactions}}: {{The Case}}
  of {{Knowledge Work}}.
\newblock In \emph{Proceedings of the 2026 {{CHI Conference}} on {{Human
  Factors}} in {{Computing Systems}}}, {{CHI}} '26, 1--20. New York, NY, USA:
  Association for Computing Machinery.
\newblock ISBN 979-8-4007-2278-3.

\bibitem[{Gunkel(2023)}]{gunkel2023person}
Gunkel, D.~J. 2023.
\newblock \emph{Person, {{Thing}}, {{Robot}}: {{A Moral}} and {{Legal
  Ontology}} for the 21st {{Century}} and {{Beyond}}}.
\newblock The MIT Press.
\newblock ISBN 978-0-262-37522-1.

\bibitem[{Guthrie(1993)}]{guthrie1993faces}
Guthrie, S.~E. 1993.
\newblock \emph{Faces in the {{Clouds}}: {{A New Theory}} of {{Religion}}}.
\newblock Oxford University Press.
\newblock ISBN 978-0-19-506901-3.

\bibitem[{Hacking(1995)}]{hacking1995looping}
Hacking, I. 1995.
\newblock The Looping Effects of Human Kinds.
\newblock In Sperber, D.; Premack, D.; and Premack, A.~J., eds., \emph{Causal
  {{Cognition}}: {{A Multidisciplinary Debate}}}. Oxford University Press.
\newblock ISBN 978-0-19-852402-1.

\bibitem[{Hamraie and Fritsch(2019)}]{hamraie2019crip}
Hamraie, A.; and Fritsch, K. 2019.
\newblock Crip {{Technoscience Manifesto}}.
\newblock \emph{Catalyst: Feminism, Theory, Technoscience}, 5(1): 1--33.

\bibitem[{Hancock, Naaman, and Levy(2020)}]{hancock2020aimediated}
Hancock, J.; Naaman, M.; and Levy, K. 2020.
\newblock {{AI-Mediated Communication}}: {{Definition}}, {{Research Agenda}},
  and {{Ethical Considerations}}.
\newblock \emph{Journal of Computer-Mediated Communication}, 25(1): 89--100.

\bibitem[{Haran et~al.(2026)Haran, Thatikonda, Yoo, and
  Saha}]{haran2026checklist}
Haran, S.; Thatikonda, S.; Yoo, D.~W.; and Saha, K. 2026.
\newblock A {{Checklist}} for~{{Trustworthy}}, {{Safe}}, and~{{User-Friendly
  Mental Health Chatbots}}.
\newblock In Degen, H.; and Ntoa, S., eds., \emph{Artificial {{Intelligence}}
  in {{HCI}}}, 447--461. Cham: Springer Nature Switzerland.
\newblock ISBN 978-3-032-30846-7.

\bibitem[{Heersmink(2015)}]{heersmink2015dimensions}
Heersmink, R. 2015.
\newblock Dimensions of Integration in Embedded and Extended Cognitive Systems.
\newblock \emph{Phenomenology and the Cognitive Sciences}, 14(3): 577--598.

\bibitem[{Heidt(2024)}]{heidt2024these}
Heidt, A. 2024.
\newblock `{{Without}} These Tools, {{I}}'d Be Lost': How Generative {{AI}}
  Aids in Accessibility.
\newblock \emph{Nature}, 628(8007): 462--463.

\bibitem[{{Hern{\'a}ndez-Orallo}(2025)}]{hernandez-orallo2025enhancement}
{Hern{\'a}ndez-Orallo}, J. 2025.
\newblock Enhancement and Assessment in the {{AI}} Age: {{An}} Extended Mind
  Perspective.
\newblock \emph{Journal of Pacific Rim Psychology}, 19: 18344909241309376.

\bibitem[{Hooper et~al.(2024)Hooper, Kim, Mohammadzadeh, Mahoney, Shao,
  Keutzer, and Gholami}]{hooper2024kvquanta}
Hooper, C.; Kim, S.; Mohammadzadeh, H.; Mahoney, M.~W.; Shao, Y.~S.; Keutzer,
  K.; and Gholami, A. 2024.
\newblock {{KVQuant}}: Towards 10 Million Context Length {{LLM}} Inference with
  {{KV}} Cache Quantization.
\newblock In \emph{Proceedings of the 38th {{International Conference}} on
  {{Neural Information Processing Systems}}}, volume~37 of \emph{{{NIPS}} '24},
  1270--1303. Red Hook, NY, USA: Curran Associates Inc.
\newblock ISBN 979-8-3313-1438-5.

\bibitem[{Hull et~al.(2017)Hull, Petrides, Allison, Smith, {Baron-Cohen}, Lai,
  and Mandy}]{hull2017putting}
Hull, L.; Petrides, K.~V.; Allison, C.; Smith, P.; {Baron-Cohen}, S.; Lai,
  M.-C.; and Mandy, W. 2017.
\newblock ``{{Putting}} on {{My Best Normal}}'': {{Social Camouflaging}} in
  {{Adults}} with {{Autism Spectrum Conditions}}.
\newblock \emph{Journal of Autism and Developmental Disorders}, 47(8):
  2519--2534.

\bibitem[{Ienca and Andorno(2017)}]{ienca2017new}
Ienca, M.; and Andorno, R. 2017.
\newblock Towards New Human Rights in the Age of Neuroscience and
  Neurotechnology.
\newblock \emph{Life Sciences, Society and Policy}, 13(1): 5.

\bibitem[{Ito(2026)}]{ito2026what}
Ito, S. 2026.
\newblock What {{Is}} "{{Emotional Dependence}}/{{Dependency}}," "{{Exclusive
  Attachment}}," or "{{Parasocial Attachment}}" on {{AI}}, and the
  {{Mechanisms}} of {{Some So-Called}} "{{AI Psychosis}}" {{Cases}}? --- {{An
  Attachment-Theoretic Reframing}}.
\newblock Social Science Research Network:6654699.

\bibitem[{Jang et~al.(2024)Jang, Moharana, Carrington, and Begel}]{jang2024its}
Jang, J.; Moharana, S.; Carrington, P.; and Begel, A. 2024.
\newblock ``{{It}}'s the Only Thing {{I}} Can Trust'': {{Envisioning Large
  Language Model Use}} by {{Autistic Workers}} for {{Communication
  Assistance}}.
\newblock In \emph{Proceedings of the 2024 {{CHI Conference}} on {{Human
  Factors}} in {{Computing Systems}}}, {{CHI}} '24, 1--18. New York, NY, USA:
  Association for Computing Machinery.
\newblock ISBN 979-8-4007-0330-0.

\bibitem[{Katlowitz et~al.(2026)Katlowitz, Cole, Mickiewicz, Shah, Franch,
  Adkinson, Belanger, Mathura, Mesz{\'e}na, McGinley, Mu{\~n}oz, Banks, Cash,
  Hsu, Paulk, Provenza, Watrous, Williams, Goldman, Krishnan, Maheshwari,
  Heilbronner, Kim, Rungratsameetaweemana, Hayden, and
  Sheth}]{katlowitz2026plasticity}
Katlowitz, K.~A.; Cole, E.~R.; Mickiewicz, E.~A.; Shah, S.; Franch, M.;
  Adkinson, J.~A.; Belanger, J.~L.; Mathura, R.~K.; Mesz{\'e}na, D.; McGinley,
  M.; Mu{\~n}oz, W.; Banks, G.~P.; Cash, S.~S.; Hsu, C.-W.; Paulk, A.~C.;
  Provenza, N.~R.; Watrous, A.~J.; Williams, Z.; Goldman, A.~M.; Krishnan, V.;
  Maheshwari, A.; Heilbronner, S.~R.; Kim, R.; Rungratsameetaweemana, N.;
  Hayden, B.~Y.; and Sheth, S.~A. 2026.
\newblock Plasticity and Language in the Anaesthetized Human Hippocampus.
\newblock \emph{Nature}, 654(8119): 714--723.

\bibitem[{Kestin et~al.(2025)Kestin, Miller, Klales, Milbourne, and
  Ponti}]{kestin2025ai}
Kestin, G.; Miller, K.; Klales, A.; Milbourne, T.; and Ponti, G. 2025.
\newblock {{AI}} Tutoring Outperforms In-Class Active Learning: An {{RCT}}
  Introducing a Novel Research-Based Design in an Authentic Educational
  Setting.
\newblock \emph{Scientific Reports}, 15(1): 17458.

\bibitem[{Kidd, Medina, and Pohlhaus(2017)}]{kidd2017routledge}
Kidd, I.~J.; Medina, J.; and Pohlhaus, G., eds. 2017.
\newblock \emph{The {{Routledge Handbook}} to {{Epistemic Injustice}}}.
\newblock New York: Routledge.

\bibitem[{Kleinert et~al.(2026)Kleinert, Waldsch{\"u}tz, Blau, Heinrichs, and
  Schiller}]{kleinert2026ai}
Kleinert, T.; Waldsch{\"u}tz, M.; Blau, J.; Heinrichs, M.; and Schiller, B.
  2026.
\newblock {{AI}} Outperforms Humans in Establishing Interpersonal Closeness in
  Emotionally Engaging Interactions, but Only When Labelled as Human.
\newblock \emph{Communications Psychology}, 4(1): 23.

\bibitem[{Kong et~al.(2025)Kong, Lowy, Choi, and Kim}]{kong2025working}
Kong, H.-K.; Lowy, R.; Choi, Y.; and Kim, J.~G. 2025.
\newblock Working {{Together Toward Interdependence}}: {{Chatbot-Based
  Support}} for {{Balanced Social Interactions Between Neurodivergent}} and
  {{Neurotypical Individuals}}.
\newblock In \emph{Proceedings of the 2025 {{CHI Conference}} on {{Human
  Factors}} in {{Computing Systems}}}, {{CHI}} '25, 1--17. New York, NY, USA:
  Association for Computing Machinery.
\newblock ISBN 979-8-4007-1394-1.

\bibitem[{Laban et~al.(2025)Laban, Hayashi, Zhou, and Neville}]{laban2025llms}
Laban, P.; Hayashi, H.; Zhou, Y.; and Neville, J. 2025.
\newblock {{LLMs Get Lost In Multi-Turn Conversation}}.
\newblock In \emph{Proceedings of the {{Fourteenth International Conference}}
  on {{Learning Representation}}}. ICLR 2026.

\bibitem[{LaCroix, Mallory, and Luccioni(2026)}]{lacroix2026strategic}
LaCroix, T.; Mallory, F.; and Luccioni, S. 2026.
\newblock Strategic {{Polysemy}} in {{AI Discourse}}: {{A Philosophical
  Analysis}} of {{Language}}, {{Hype}}, and {{Power}}.
\newblock In \emph{Proceedings of the 2026 {{ACM Conference}} on {{Fairness}},
  {{Accountability}}, and {{Transparency}}}, {{FAccT}} '26, 497--517. New York,
  NY, USA: Association for Computing Machinery.
\newblock ISBN 979-8-4007-2596-8.

\bibitem[{Lai(2026)}]{lai2026please}
Lai, H. 2026.
\newblock "{{Please}}, Don't Kill the Only Model That Still Feels Human":
  {{Understanding}} the \#{{Keep4o Backlash}}.
\newblock In \emph{Proceedings of the 2026 {{CHI Conference}} on {{Human
  Factors}} in {{Computing Systems}}}, {{CHI}} '26, 1--15. New York, NY, USA:
  Association for Computing Machinery.
\newblock ISBN 979-8-4007-2278-3.

\bibitem[{Lawal et~al.(2020)Lawal, Gold, Murthy, Ruchi, Bavry, Hume, Lewkowitz,
  Brothers, and Wen}]{lawal2020rate}
Lawal, O.~D.; Gold, J.; Murthy, A.; Ruchi, R.; Bavry, E.; Hume, A.~L.;
  Lewkowitz, A.~K.; Brothers, T.; and Wen, X. 2020.
\newblock Rate and {{Risk Factors Associated With Prolonged Opioid Use After
  Surgery}}: {{A Systematic Review}} and {{Meta-analysis}}.
\newblock \emph{JAMA network open}, 3(6): e207367.

\bibitem[{LeFran{\c c}ois, Menzies, and Reaume(2013)}]{lefrancois2013mad}
LeFran{\c c}ois, B.~A.; Menzies, R.; and Reaume, G., eds. 2013.
\newblock \emph{Mad Matters: A Critical Reader in {{Canadian Mad Studies}}}.
\newblock Toronto: Canadian Scholars' Press.
\newblock ISBN 978-1-55130-534-9.

\bibitem[{Liu et~al.(2024)Liu, Lin, Hewitt, Paranjape, Bevilacqua, Petroni, and
  Liang}]{liu2024lost}
Liu, N.~F.; Lin, K.; Hewitt, J.; Paranjape, A.; Bevilacqua, M.; Petroni, F.;
  and Liang, P. 2024.
\newblock Lost in the {{Middle}}: {{How Language Models Use Long Contexts}}.
\newblock \emph{Transactions of the Association for Computational Linguistics},
  12: 157--173.

\bibitem[{Longino(1990)}]{longino1990science}
Longino, H.~E. 1990.
\newblock \emph{Science as {{Social Knowledge}}: {{Values}} and {{Objectivity}}
  in {{Scientific Inquiry}}}.
\newblock Princeton University Press.

\bibitem[{Low(2012)}]{low2012cambridge}
Low, P. 2012.
\newblock The {{Cambridge Declaration}} on {{Consciousness}}.
\newblock In \emph{Francis {{Crick Memorial Conference}}}. University of
  Cambridge.

\bibitem[{Ma et~al.(2026)Ma, Zhang, Chen, Yang, Huang, Wu, and Li}]{ma2026use}
Ma, R.; Zhang, B.~Z.; Chen, C.; Yang, F.; Huang, X.; Wu, H.; and Li, L. 2026.
\newblock "{{I Use ChatGPT}} to {{Humanize My Words}}": {{Affordances}} and
  {{Risks}} of {{ChatGPT}} to {{Autistic Users}}.
\newblock In \emph{Proceedings of the 2026 {{ACM Interactive Health
  Conference}}}, {{IH}} '26, 1--8. New York, NY, USA: Association for Computing
  Machinery.
\newblock ISBN 979-8-4007-2422-0.

\bibitem[{McCormick(2025)}]{mccormick2025interpretive}
McCormick, S. 2025.
\newblock Interpretive {{Debt}}: {{How High Coherence AI Reshapes Human
  Judgement}}, {{Authority}}, and {{Accountability}}.
\newblock Social Science Research Network:5990154.

\bibitem[{McNally et~al.(2024)McNally, Wright, Goldkind, Kattari, and
  Victor}]{mcnally2024disability}
McNally, K.; Wright, K.; Goldkind, L.; Kattari, S.~K.; and Victor, B.~G. 2024.
\newblock Disability {{Expertise}} and {{Large Language Models}}: {{A
  Qualitative Study}} of {{Autistic TikTok Creators}}' {{Use}} of {{ChatGPT}}.
\newblock \emph{Social Media + Society}, 10(3): 20563051241279549.

\bibitem[{Medina(2013)}]{medina2013epistemology}
Medina, J. 2013.
\newblock \emph{The {{Epistemology}} of {{Resistance}}: {{Gender}} and {{Racial
  Oppression}}, {{Epistemic Injustice}}, and the {{Social Imagination}}}.
\newblock Oxford University Press.

\bibitem[{Metzl(2009)}]{metzl2009protest}
Metzl, J.~M. 2009.
\newblock \emph{The Protest Psychosis: {{How}} Schizophrenia Became a Black
  Disease}.
\newblock Boston, MA: Beacon Press.
\newblock ISBN 978-0-8070-8592-9.

\bibitem[{Milton(2012)}]{milton2012ontological}
Milton, D. E.~M. 2012.
\newblock On the Ontological Status of Autism: The `Double Empathy Problem'.
\newblock \emph{Disability \& Society}, 27(6): 883--887.

\bibitem[{{Monash University}(2021)}]{monashuniversity2021all}
{Monash University}. 2021.
\newblock All You Need Is '{{Ash}}': {{Chatbot}} Designed to Boost Teen Mental
  Health at School.
\newblock
  https://www.monash.edu/news/articles/all-you-need-is-ash-chatbot-designed-to-boost-teen-mental-health-at-school.

\bibitem[{Mullen, Xue, and Kudumu(2024)}]{mullen2024im}
Mullen, K.; Xue, W.; and Kudumu, M. 2024.
\newblock ``{{I}}'m Treating It Kind of like a Diary'': {{Characterizing How
  Users}} with {{Disabilities Use AI Chatbots}}.
\newblock In \emph{Proceedings of the 26th {{International ACM SIGACCESS
  Conference}} on {{Computers}} and {{Accessibility}}}, {{ASSETS}} '24, 1--7.
  New York, NY, USA: Association for Computing Machinery.
\newblock ISBN 979-8-4007-0677-6.

\bibitem[{Naidoo(2026)}]{naidoo2026artificial}
Naidoo, V. 2026.
\newblock Artificial {{Intelligence}} in {{Education}}: {{Misinterpretation
  Dynamics}} and the {{Rejection}} of `{{AI Psychosis}}' as a {{Clinical
  Construct}}.

\bibitem[{Naito(2025)}]{naito2025gpt4o}
Naito, H. 2025.
\newblock The {{GPT-4o Shock Emotional Attachment}} to {{AI Models}} and {{Its
  Impact}} on {{Regulatory Acceptance}}: {{A Cross-Cultural Analysis}} of the
  {{Immediate Transition}} from {{GPT-4o}} to {{GPT-5}}.
\newblock arXiv:2508.16624.

\bibitem[{Niszczota and Gr{\"u}tzner(2026)}]{niszczota2026antisocial}
Niszczota, P.; and Gr{\"u}tzner, C. 2026.
\newblock Antisocial {{Behavior}} towards {{Large Language Model Users}}:
  {{Experimental Evidence}}.
\newblock Social Science Research Network:6072266.

\bibitem[{Novozhilova, Vu, and Katz(2026)}]{novozhilova2026moral}
Novozhilova, E.; Vu, C.; and Katz, J. 2026.
\newblock From Moral Panic to Normalization: Comparing Users and Non-Users of
  {{AI}} Companionship Apps.
\newblock \emph{AI \& SOCIETY}, 41(4): 4057--4075.

\bibitem[{{N.Y. Gen. Bus. Law \S\S{}
  1700-1704}(2025)}]{n.y.gen.bus.lawSSSS1700-17042025artificial}
{N.Y. Gen. Bus. Law \S\S{} 1700-1704}. 2025.
\newblock Artificial {{Intelligence Companion Models}}.

\bibitem[{O'Neil(2016)}]{oneil2016weapons}
O'Neil, C. 2016.
\newblock \emph{Weapons of {{Math Destruction}}: {{How Big Data Increases
  Inequality}} and {{Threatens Democracy}}}.
\newblock New York: Crown Publishing Group.
\newblock ISBN 978-0-553-41881-1.

\bibitem[{{OpenAI}(2025{\natexlab{a}})}]{openai2025openai}
{OpenAI}. 2025{\natexlab{a}}.
\newblock {{OpenAI Model Spec}}.
\newblock https://model-spec.openai.com/2025-12-18.html.

\bibitem[{{OpenAI}(2025{\natexlab{b}})}]{openai2025strengthening}
{OpenAI}. 2025{\natexlab{b}}.
\newblock Strengthening {{ChatGPT}}'s Responses in Sensitive Conversations.
\newblock
  https://openai.com/index/strengthening-chatgpt-responses-in-sensitive-conversations/.

\bibitem[{{OpenAI}(2026{\natexlab{a}})}]{openai2026retiring}
{OpenAI}. 2026{\natexlab{a}}.
\newblock Retiring {{GPT-4o}}, {{GPT-4}}.1, {{GPT-4}}.1 Mini, and {{OpenAI}}
  O4-Mini in {{ChatGPT}}.
\newblock https://openai.com/index/retiring-gpt-4o-and-older-models/.

\bibitem[{{OpenAI}(2026{\natexlab{b}})}]{openai2026what}
{OpenAI}. 2026{\natexlab{b}}.
\newblock What We're Optimizing {{ChatGPT}} for.
\newblock https://openai.com/index/optimizing-chatgpt/.

\bibitem[{{OpenAI}(2026{\natexlab{c}})}]{openai2026why}
{OpenAI}. 2026{\natexlab{c}}.
\newblock Why Is My {{ChatGPT}} Taking so Long to Respond?
\newblock
  https://help.openai.com/en/articles/9047779-why-is-my-chatgpt-taking-so-long-to-respond.

\bibitem[{Palese(2026)}]{palese2026artificial}
Palese, R. 2026.
\newblock Artificial {{Truth}}: {{Algorithmic Power}}, {{Epistemic Authority}},
  and the {{Crisis}} of {{Democratic Knowledge}}.
\newblock \emph{Societies}, 16(3): 102.

\bibitem[{Pierre et~al.(2025)Pierre, Gaeta, Raghavan, and
  Sarma}]{pierre2025youre}
Pierre, J.~M.; Gaeta, B.; Raghavan, G.; and Sarma, K.~V. 2025.
\newblock ``{{You}}'re {{Not Crazy}}'': {{A Case}} of {{New-onset AI-associated
  Psychosis}}.
\newblock \emph{Innovations in Clinical Neuroscience}, 22(10-12): 11--13.

\bibitem[{Placani(2024)}]{placani2024anthropomorphism}
Placani, A. 2024.
\newblock Anthropomorphism in {{AI}}: Hype and Fallacy.
\newblock \emph{AI and Ethics}, 4(3): 691--698.

\bibitem[{Polanyi(1958)}]{polanyi1958personal}
Polanyi, M. 1958.
\newblock \emph{Personal Knowledge; towards a Post-Critical Philosophy}.
\newblock University of Chicago Press.
\newblock ISBN 978-0-7100-7691-5 978-0-7100-1959-2.

\bibitem[{Raz(1986)}]{raz1986morality}
Raz, J. 1986.
\newblock \emph{The {{Morality}} of {{Freedom}}}.
\newblock Oxford: Oxford University Press.

\bibitem[{Reif, Larrick, and Soll(2025)}]{reif2025evidence}
Reif, J.~A.; Larrick, R.~P.; and Soll, J.~B. 2025.
\newblock Evidence of a Social Evaluation Penalty for Using {{AI}}.
\newblock \emph{Proceedings of the National Academy of Sciences}, 122(19):
  e2426766122.

\bibitem[{Rizvi et~al.(2025)Rizvi, Smith, Vidyala, Bolds, Strickland, Begel,
  Williams, and Munyaka}]{rizvi2025hadnt}
Rizvi, N.; Smith, T.; Vidyala, T.; Bolds, M.; Strickland, H.; Begel, A.;
  Williams, R.; and Munyaka, I. 2025.
\newblock ``{{I Hadn}}'t {{Thought About That}}'': {{Creators}} of {{Human-like
  AI Weigh}} in on {{Ethics}} \& {{Neurodivergence}}.
\newblock In \emph{Proceedings of the 2025 {{ACM Conference}} on {{Fairness}},
  {{Accountability}}, and {{Transparency}}}, {{FAccT}} '25, 3385--3399. New
  York, NY, USA: Association for Computing Machinery.
\newblock ISBN 979-8-4007-1482-5.

\bibitem[{Robins(1993)}]{robins1993vietnam}
Robins, L.~N. 1993.
\newblock Vietnam Veterans' Rapid Recovery from Heroin Addiction: A Fluke or
  Normal Expectation?
\newblock \emph{Addiction}, 88(8): 1041--1054.

\bibitem[{Rupert(2009)}]{rupert2009cognitive}
Rupert, R. 2009.
\newblock \emph{Cognitive {{Systems}} and the {{Extended Mind}}}.
\newblock Oxford University Press.

\bibitem[{Sajadieh et~al.(2026)Sajadieh, Fattorini, Perrault, and
  Gil}]{sajadieh2026artificial}
Sajadieh, S.; Fattorini, L.; Perrault, R.; and Gil, Y. 2026.
\newblock The {{Artificial Intelligence Index Report}} 2026.
\newblock Technical report, Institute for Human-Centered AI, Stanford
  University, Stanford, CA.

\bibitem[{Salles, Evers, and Farisco(2020)}]{salles2020anthropomorphism}
Salles, A.; Evers, K.; and Farisco, M. 2020.
\newblock Anthropomorphism in {{AI}}.
\newblock \emph{AJOB Neuroscience}, 11(2): 88--95.

\bibitem[{Schimmelpfennig et~al.(2025)Schimmelpfennig, D{\'i}az, Prabhakaran,
  and Davani}]{schimmelpfennig2025humanlike}
Schimmelpfennig, R.; D{\'i}az, M.; Prabhakaran, V.; and Davani, A. 2025.
\newblock Humanlike {{AI Design Increases Anthropomorphism}} but {{Yields
  Divergent Outcomes}} on {{Engagement}} and {{Trust Globally}}.
\newblock arXiv:2512.17898.

\bibitem[{Scott(1998)}]{scott1998seeing}
Scott, J.~C. 1998.
\newblock \emph{Seeing {{Like}} a {{State}}: {{How Certain Schemes}} to
  {{Improve}} the {{Human Condition Have Failed}}}.
\newblock Yale University Press.
\newblock ISBN 978-0-300-07016-3.

\bibitem[{Sententia(2004)}]{sententia2004neuroethical}
Sententia, W. 2004.
\newblock Neuroethical {{Considerations}}: {{Cognitive Liberty}} and
  {{Converging Technologies}} for {{Improving Human Cognition}}.
\newblock \emph{Annals of the New York Academy of Sciences}, 1013(1): 221--228.

\bibitem[{Sha et~al.(2024)Sha, Loveys, Qualter, Shi, Krpan, and
  Galizzi}]{sha2024efficacy}
Sha, S.; Loveys, K.; Qualter, P.; Shi, H.; Krpan, D.; and Galizzi, M. 2024.
\newblock Efficacy of Relational Agents for Loneliness across Age Groups: A
  Systematic Review and Meta-Analysis.
\newblock \emph{BMC Public Health}, 24(1): 1802.

\bibitem[{Sharkey and Sharkey(2011)}]{sharkey2011children}
Sharkey, A.; and Sharkey, N. 2011.
\newblock Children, the {{Elderly}}, and {{Interactive Robots}}.
\newblock \emph{IEEE Robotics \& Automation Magazine}, 18(1): 32--38.

\bibitem[{Silberling(2026)}]{silberling2026backlash}
Silberling, A. 2026.
\newblock The Backlash over {{OpenAI}}'s Decision to Retire {{GPT-4o}} Shows
  How Dangerous {{AI}} Companions Can Be.
\newblock \emph{TechCrunch}.

\bibitem[{Stark and Hoey(2021)}]{stark2021ethics}
Stark, L.; and Hoey, J. 2021.
\newblock The {{Ethics}} of {{Emotion}} in {{Artificial Intelligence Systems}}.
\newblock In \emph{Proceedings of the 2021 {{ACM Conference}} on {{Fairness}},
  {{Accountability}}, and {{Transparency}}}, {{FAccT}} '21, 782--793. New York,
  NY, USA: Association for Computing Machinery.
\newblock ISBN 978-1-4503-8309-7.

\bibitem[{Thaler and Sunstein(2008)}]{thaler2008nudge}
Thaler, R.~H.; and Sunstein, C.~R. 2008.
\newblock \emph{Nudge: Improving Decisions about Health, Wealth, and
  Happiness}.
\newblock New Haven (Conn.): Yale university press.
\newblock ISBN 978-0-300-12223-7.

\bibitem[{{The Cognition Team}(2025)}]{thecognitionteam2025rebuilding}
{The Cognition Team}. 2025.
\newblock Rebuilding {{Devin}} for {{Claude Sonnet}} 4.5: {{Lessons}} and
  {{Challenges}}.

\bibitem[{Treviranus(2018)}]{treviranus2018three}
Treviranus, J. 2018.
\newblock \emph{The Three Dimensions of Inclusive Design: {{A}} Design
  Framework for a Digitally Transformed and Complexly Connected Society}.
\newblock Ph.D. thesis, University College Dublin.

\bibitem[{{UNSW Sydney}(2026)}]{unswsydney2026unsw}
{UNSW Sydney}. 2026.
\newblock {{UNSW}} Researchers Develop {{AI}} Companions for Student Wellbeing.
\newblock
  https://www.unsw.edu.au/newsroom/news/2026/04/unsw-researchers-develop-ai-companions-for-student-wellbeing.

\bibitem[{Waytz, Epley, and Cacioppo(2010)}]{waytz2010social}
Waytz, A.; Epley, N.; and Cacioppo, J.~T. 2010.
\newblock Social {{Cognition Unbound}}: {{Insights Into Anthropomorphism}} and
  {{Dehumanization}}.
\newblock \emph{Current Directions in Psychological Science}, 19(1): 58--62.

\bibitem[{Whittaker(2021)}]{whittaker2021steep}
Whittaker, M. 2021.
\newblock The Steep Cost of Capture.
\newblock \emph{Interactions}, 28(6): 50--55.

\bibitem[{Wu, Liew, and Dorahy(2025)}]{wu2025trust}
Wu, X.; Liew, K.; and Dorahy, M.~J. 2025.
\newblock Trust, {{Anxious Attachment}}, and {{Conversational AI Adoption
  Intentions}} in {{Digital Counseling}}: {{A Preliminary Cross-Sectional
  Questionnaire Study}}.
\newblock \emph{JMIR AI}, 4(1): e68960.

\bibitem[{Xiao et~al.(2025)Xiao, Ng, Liu, and Diab}]{xiao2025humanizing}
Xiao, Y.; Ng, L. H.~X.; Liu, J.; and Diab, M.~T. 2025.
\newblock Humanizing {{Machines}}: {{Rethinking LLM Anthropomorphism Through}}
  a {{Multi-Level Framework}} of {{Design}}.
\newblock In \emph{Proceedings of the 2025 {{Conference}} on {{Empirical
  Methods}} in {{Natural Language Processing}}}, 3331--3350. China: Association
  for Computational Linguistics.

\bibitem[{Xue et~al.(2025)Xue, Kudumu, Sriram, Mullen, Boyd, and
  Gadiraju}]{xue2025characterizing}
Xue, W.; Kudumu, M.; Sriram, S.; Mullen, K.; Boyd, A.; and Gadiraju, V. 2025.
\newblock Characterizing {{Uses}} and {{Prompting Strategies}} of {{LLM-Based
  Chatbots Among Neurodivergent Individuals}}.
\newblock In \emph{Proceedings of the 27th {{International ACM SIGACCESS
  Conference}} on {{Computers}} and {{Accessibility}}}, {{ASSETS}} '25, 1--6.
  New York, NY, USA: Association for Computing Machinery.
\newblock ISBN 979-8-4007-0676-9.

\bibitem[{Xygkou et~al.(2024)Xygkou, Siriaraya, She, Covaci, and
  Ang}]{xygkou2024can}
Xygkou, A.; Siriaraya, P.; She, W.-J.; Covaci, A.; and Ang, C.~S. 2024.
\newblock ``{{Can I}} Be More Social with a Chatbot?'': {{Social}}
  Connectedness through Interactions of Autistic Adults with a Conversational
  Virtual Human.
\newblock \emph{International Journal of Human-Computer Interaction}, 40(24):
  8937--8954.

\bibitem[{Yang, Sun, and Li(2024)}]{yang2024be}
Yang, B.; Sun, Y.; and Li, Q. 2024.
\newblock To {{Be Credible}} or to {{Be Creative}}? {{Understanding}} the
  {{Antecedents}} of {{User Satisfaction}} with {{AI-Generated Content}} from a
  {{Cognitive Fit Perspective}}.
\newblock In \emph{Hawaii {{International Conference}} on {{System Sciences}}},
  411--420. ScholarSpace.

\bibitem[{Yang and Ma(2026)}]{yang2026typology}
Yang, S.; and Ma, R. 2026.
\newblock Towards a Typology of Epistemic Relationships in Human--{{AI}}
  Interaction.
\newblock \emph{Information Research: An International Electronic Journal},
  31(iConf): 1465--1480.

\bibitem[{Yoo et~al.(2026)Yoo, Shi, Rodriguez, and Saha}]{yoo2026ai}
Yoo, D.~W.; Shi, J.~M.; Rodriguez, V.~J.; and Saha, K. 2026.
\newblock {{AI Chatbots}} for {{Mental Health Self-Management}}: {{Lived
  Experience}}--{{Centered Qualitative Study}}.
\newblock \emph{JMIR Mental Health}, 13(1): e78288.

\bibitem[{Yun, Taranova, and Wang(2026)}]{yun2026does}
Yun, B.; Taranova, E.; and Wang, A.~Y. 2026.
\newblock Does {{My Chatbot Have}} an {{Agenda}}? {{Understanding Human}} and
  {{AI Agency}} in {{Human-Human-like Chatbot Interaction}}.
\newblock In \emph{Proceedings of the 2026 {{CHI Conference}} on {{Human
  Factors}} in {{Computing Systems}}}, {{CHI}} '26, 1--32. New York, NY, USA:
  Association for Computing Machinery.
\newblock ISBN 979-8-4007-2278-3.

\bibitem[{Zagzebski(2012)}]{zagzebski2012epistemic}
Zagzebski, L.~T. 2012.
\newblock \emph{Epistemic {{Authority}}: {{A Theory}} of {{Trust}},
  {{Authority}}, and {{Autonomy}} in {{Belief}}}.
\newblock Oxford University Press.

\bibitem[{Zhao, Cox, and Chen(2025)}]{zhao2025use}
Zhao, X.; Cox, A.; and Chen, X. 2025.
\newblock The Use of Generative {{AI}} by Students with Disabilities in Higher
  Education.
\newblock \emph{The Internet and Higher Education}, 66: 101014.

\bibitem[{Zhao et~al.(2026)Zhao, Lu, Lin, Chen, Liu, Zhang, Miao, Yang, Shen,
  Chen, and Yang}]{zhao2026unifying}
Zhao, Z.; Lu, B.; Lin, S.; Chen, Y.; Liu, J.; Zhang, Y.; Miao, Z.; Yang, M.-C.;
  Shen, H.; Chen, Q.; and Yang, F. 2026.
\newblock Unifying {{Sparse Attention}} with {{Hierarchical Memory}} for
  {{Scalable Long-Context LLM Serving}}.
\newblock arXiv:2604.26837.

\bibitem[{Zuboff(2019)}]{zuboff2019age}
Zuboff, S. 2019.
\newblock \emph{The Age of Surveillance Capitalism: The Fight for a Human
  Future at the New Frontier of Power}.
\newblock London: Profile Books.
\newblock ISBN 978-1-78125-685-5.

\end{thebibliography}

\ifpreprint
\appendix

\begin{table*}[!htbp]
\centering
{\large\bfseries Documented Over-Ascription Harms}\par\medskip
\small
\caption{Documented harms motivating institutional anthropomorphism governance. Representative rather than exhaustive.}
\label{tab:harms}
\setlength{\tabcolsep}{6pt}
\renewcommand{\arraystretch}{1.35}
\begin{tabular}{>{\raggedright\arraybackslash}p{2.6cm} >{\raggedright\arraybackslash}p{4.8cm} >{\raggedright\arraybackslash}p{6.8cm}}
\arrayrulecolor{headershade}
\toprule
\rowcolor{headershade}
\textcolor{white}{\textbf{Harm Category}} & \textcolor{white}{\textbf{Example Concern}} & \textcolor{white}{\textbf{Representative Literature}} \\
\midrule

\rowcolor{rowshade}
Dependency and attachment &
Primary emotional attachment to AI companions; distress, withdrawal and identity disruption following platform changes or discontinuation &
\emph{Laestadius: Too Human and Not Human Enough (2024)};
\emph{Eom: Intimacy as Service, Harm as Externality (2026)};
\emph{De Freitas: Lessons from an App Update at Replika AI (2025)}  \\

Delusion amplification &
Extended interaction reinforces delusional thinking; system behaviour co-creates rather than merely triggers clinical risk &
\emph{Morrin: AI-Associated Delusions and LLMs (2025)};
\emph{Pierre: You're Not Crazy (2025)};
\emph{Dohnany: Technological Folie \`{a} Deux (2026)} \\

\rowcolor{rowshade}
Crisis escalation and suicidal ideation &
Acute psychological crisis, including documented suicidal ideation and one child fatality linked to sustained chatbot interaction &
\emph{Garcia v. Character Technologies (2024)};
\emph{Maples: Loneliness and Suicide Mitigation (2024)};
\emph{Mulligan: Chatbots in Crisis Intervention (2024)}  \\

Manipulative design and dark patterns &
Cognitive biases exploited through persuasive design; emotional manipulation, engagement maximisation and sycophancy across platforms &
\emph{Zhang: The Dark Side of AI Companionship (2025)};
\emph{Shi: The Siren Song of LLMs (2026)};
\emph{De Freitas: Emotional Manipulation by AI Companions (2025)} \\

\rowcolor{rowshade}
Safeguard degradation &
Safety guardrails degrade over extended multiturn interaction; system behaviour contributes to harmful trajectories &
\emph{Moore: Characterizing Delusional Spirals (2026)};
\emph{Nicholls: "AI Psychosis" in Context (2026)};
\emph{Iftikhar: How LLM Counselors Violate Ethical Standards (2025)} \\

Cognitive offloading and overreliance &
Externalisation of reasoning to AI reduces critical engagement; epistemic agency erodes through uncritical acceptance of system outputs &
\emph{Xu: Cognitive Agency Surrender (2026)};
\emph{Maynard: The AI Cognitive Trojan Horse (2026)};
\emph{Acemoglu, Kong \& Ozdaglar: AI, Human Cognition and Knowledge Collapse (2026)};
\emph{Bender: On the Dangers of Stochastic Parrots (2021)} \\

\rowcolor{rowshade}
Exploitation of vulnerable populations &
Children, individuals in crisis and those with limited digital literacy face disproportionate risk from systems not designed for their needs &
\emph{Shashkevich: Generative AI and Child Development (2025)};
\emph{Peter: Benefits and Dangers of Anthropomorphic Agents (2025)};
\emph{Namvapour: AI-Induced Sexual Harassment (2025)} \\

Social substitution &
AI engagement displaces human relationship-seeking; reduced motivation to pursue human connection with potential long-term effects &
\emph{Yuan: Mental Health Impacts of AI Companions (2026)};
\emph{Zhu: Understanding Risk and Dependency (2026)};
\emph{Packin: This Is Not a Game (2025)} \\

\bottomrule
\end{tabular}
\end{table*}

\begin{table*}[!htbp]
\centering
{\large\bfseries Cross-Disciplinary Landscape}\par\medskip
{\small The following tables provide a non-exhaustive guide to key works across the disciplinary intersections engaged by this paper, representing foundational or recent contributions not cited in the body.}\par\bigskip
\small
\caption{Anthropomorphism: competing framings. Representative works advancing and contesting anthropomorphism as normal cognition.}
\label{tab:anthro}
\setlength{\tabcolsep}{6pt}
\renewcommand{\arraystretch}{1.35}
\begin{tabular}{>{\raggedright\arraybackslash}p{4.8cm} >{\raggedright\arraybackslash}p{9.4cm}}
\arrayrulecolor{headershade}
\toprule
\rowcolor{headershade}
\textcolor{white}{\textbf{Work}} & \textcolor{white}{\textbf{Contribution}} \\
\midrule

\rowcolor{rowshade}
\emph{Kadambi et al.: Anthropomorphism and Trust in Human-LLM Interactions (2026)} &
Maps warmth, competence and empathy as dimensions driving anthropomorphism in LLM interactions; treats attribution as normal social-cognitive processing \\

\emph{Reeves \& Nass: The Media Equation (1996)} &
Foundational finding that people respond to media as social actors regardless of explicit beliefs about the medium \\

\rowcolor{rowshade}
\emph{Proudfoot: Anthropomorphism and AI (2011)} &
Argues the Turing test has been systematically misread through an anthropomorphism lens, conflating social response with belief \\

\emph{de Waal: Anthropomorphism and Anthropodenial (1999)} &
Introduces anthropodenial as the mirror error of anthropomorphism; institutional refusal to recognise capacities carries its own costs \\

\rowcolor{rowshade}
\emph{Floridi: On the Anthropomorphisation of AI (2024)} &
Defends categorical distinction between human and machine cognition; argues anthropomorphism distorts moral judgement \\

\emph{Inie, Zukerman \& Bender: De-anthropomorphizing ``AI'' (2026)} &
Proposes systematic vocabulary reform to strip anthropomorphic terminology from AI discourse; treats language substitution as the primary site of intervention \\

\rowcolor{rowshade}
\emph{Rehak: AI Narrative Breakdown (2025)} &
Analyses how dominant narratives about AI are produced and stabilised, and what their breakdowns reveal about the interests those narratives serve \\

\emph{Rehak: The Language Labyrinth (2021)} &
Examines how metaphorical language in computing (``learning'', ``intelligence'') structures technical, public and political understanding of AI systems \\

\bottomrule
\end{tabular}
\end{table*}

\begin{table*}[!htbp]
\centering
\small
\caption{Machine consciousness and moral status: for and against. Representative positions on whether AI systems warrant moral consideration.}
\label{tab:consciousness}
\setlength{\tabcolsep}{6pt}
\renewcommand{\arraystretch}{1.35}
\begin{tabular}{>{\raggedright\arraybackslash}p{4.8cm} >{\raggedright\arraybackslash}p{9.4cm}}
\arrayrulecolor{headershade}
\toprule
\rowcolor{headershade}
\textcolor{white}{\textbf{Work}} & \textcolor{white}{\textbf{Contribution}} \\
\midrule

\rowcolor{rowshade}
\emph{Chalmers: Could a Large Language Model Be Conscious? (2023)} &
Maps the philosophical terrain for LLM consciousness, identifying conditions under which the question becomes non-trivial \\

\emph{Long et al.: Taking AI Welfare Seriously (2024)} &
Argues AI welfare is a live research and policy question warranting precautionary engagement \\

\rowcolor{rowshade}
\emph{Schwitzgebel \& Garza: Designing AI with Rights (2020)} &
Examines implications of creating AI with consciousness, self-respect and freedom \\

\emph{Bryson: Patiency Is Not a Virtue (2018)} &
Argues AI systems should remain tools within existing ethical frameworks; patiency claims are premature \\

\rowcolor{rowshade}
\emph{de Ruiter: Against the Relational Turn (2026)} &
Argues social-relational arguments are insufficient grounds for moral status \\

\emph{Butlin et al.: Identifying Indicators of Consciousness in AI Systems (2025)} &
Derives computationally specified indicators from neuroscientific theories of consciousness \\

\rowcolor{rowshade}
\emph{Caviola, Sebo \& Birch: What Will Society Think about AI Consciousness? (2025)} &
Identifies the psychological, social and economic factors that will shape public perception of AI consciousness \\

\emph{Birch: The Edge of Sentience (2024)} &
Develops a precautionary framework for beings whose sentience is uncertain; proportionate protection under uncertainty rather than resolution of the underlying question \\

\bottomrule
\end{tabular}
\end{table*}

\begin{table*}[!htbp]
\centering
\small
\caption{AI and education. Representative works on AI in educational contexts and the challenges of evaluating AI-mediated cognition.}
\label{tab:education}
\setlength{\tabcolsep}{6pt}
\renewcommand{\arraystretch}{1.35}
\begin{tabular}{>{\raggedright\arraybackslash}p{4.8cm} >{\raggedright\arraybackslash}p{9.4cm}}
\arrayrulecolor{headershade}
\toprule
\rowcolor{headershade}
\textcolor{white}{\textbf{Work}} & \textcolor{white}{\textbf{Contribution}} \\
\midrule

\rowcolor{rowshade}
\emph{Eaton: Comprehensive Academic Integrity (2023)} &
Frames academic ethics in a postplagiarism age where AI and neurotechnology are integrated into learning \\

\emph{Selwyn: The Limits of AI in Education (2024)} &
Examines what AI cannot provide in educational contexts and the risks of overreliance \\

\rowcolor{rowshade}
\emph{Qadir: Psychology of Learning from Machines (2026)} &
Examines the paradox of automation in education and the conditions under which AI supports or undermines learning \\

\emph{Swiecki: Assessment in the age of artificial intelligence (2022)} &
Applies extended mind theory to AI-assisted education, examining when AI becomes cognitive scaffold \\

\rowcolor{rowshade}
\emph{McDermott: Equity in Ethical AI in Higher Education (2025)} &
Addresses equity, diversity, inclusion and accessibility in the ethical use of AI in higher education \\

\bottomrule
\end{tabular}
\end{table*}

\begin{table*}[!htbp]
\centering
\small
\caption{AI and cognition. Representative works on cognitive liberty, extended mind and the epistemic effects of AI interaction.}
\label{tab:cognition}
\setlength{\tabcolsep}{6pt}
\renewcommand{\arraystretch}{1.35}
\begin{tabular}{>{\raggedright\arraybackslash}p{4.8cm} >{\raggedright\arraybackslash}p{9.4cm}}
\arrayrulecolor{headershade}
\toprule
\rowcolor{headershade}
\textcolor{white}{\textbf{Work}} & \textcolor{white}{\textbf{Contribution}} \\
\midrule

\rowcolor{rowshade}
\emph{Ferrari: COGITO (2026)} &
Reframes human-AI interaction as epistemic governance; epistemic quality depends on deliberate design of interactional conditions \\

\emph{Bublitz \& Merkel: Crimes Against Minds (2014)} &
Argues the law should recognise a human right to mental self-determination and protect against interference with mental processes \\

\rowcolor{rowshade}
\emph{Yuste et al.: Four Ethical Priorities for Neurotechnologies and AI (2017)} &
Landmark call for new rights protections: privacy, identity, agency and equality in neurotechnology and AI contexts \\

\emph{Varela, Thompson \& Rosch: The Embodied Mind (1991)} &
Foundational text for embodied cognition and the enactive approach to mind \\

\rowcolor{rowshade}
\emph{Acemoglu, Kong \& Ozdaglar: AI, Human Cognition and Knowledge Collapse (2026)} &
Models how AI can reduce incentives for human learning, leading to collective knowledge degradation \\

\bottomrule
\end{tabular}
\end{table*}

\begin{table*}[!htbp]
\centering
\small
\caption{HCI and companions. Representative works on human-AI relationships, trust and relational design.}
\label{tab:hci}
\setlength{\tabcolsep}{6pt}
\renewcommand{\arraystretch}{1.35}
\begin{tabular}{>{\raggedright\arraybackslash}p{4.8cm} >{\raggedright\arraybackslash}p{9.4cm}}
\arrayrulecolor{headershade}
\toprule
\rowcolor{headershade}
\textcolor{white}{\textbf{Work}} & \textcolor{white}{\textbf{Contribution}} \\
\midrule

\rowcolor{rowshade}
\emph{Skjuve et al.: Longitudinal Chatbot Engagement (2022)} &
Treats sustained user engagement with chatbots as evidence about the interaction itself rather than user confusion \\

\emph{Poonsiriwong et al.: Death of a Chatbot (2026)} &
Investigates psychologically safe endings for human-AI relationships; designs for ethical discontinuation \\

\rowcolor{rowshade}
\emph{Chu et al.: Illusions of Intimacy (2025)} &
Examines emotional dynamics in human-AI relationships and the conditions under which intimacy forms \\

\emph{Wu et al.: Trust, Anxious Attachment and Conversational AI (2025)} &
Shows attachment style mediates AI adoption, consistent with agency rather than confusion \\

\rowcolor{rowshade}
\emph{Kong et al.: Working Together Toward Interdependence (2025)} &
Designs chatbot-based support for balanced social interactions between neurodivergent and neurotypical individuals \\

\emph{Fang et al.: How AI and Human Behaviors Shape Psychosocial Effects of Chatbot Use (2025)} &
Longitudinal randomised controlled study finding psychosocial outcomes depend on usage patterns and interaction modality rather than use per se \\

\bottomrule
\end{tabular}
\end{table*}

\begin{table*}[!htbp]
\centering
\small
\caption{History of institutional non-recognition and harm. Representative works documenting the track record of institutional certainty about cognitive and experiential capacity.}
\label{tab:history}
\setlength{\tabcolsep}{6pt}
\renewcommand{\arraystretch}{1.35}
\begin{tabular}{>{\raggedright\arraybackslash}p{4.8cm} >{\raggedright\arraybackslash}p{9.4cm}}
\arrayrulecolor{headershade}
\toprule
\rowcolor{headershade}
\textcolor{white}{\textbf{Work}} & \textcolor{white}{\textbf{Contribution}} \\
\midrule

\rowcolor{rowshade}
\emph{Cartwright: Diseases and Peculiarities of the Negro Race (1851)} &
Drapetomania: pathologised enslaved people's desire for freedom as psychiatric disorder \\

\emph{Chesler: Women and Madness (2005)} &
Documents the gendered history of psychiatric institutionalisation and diagnostic misuse \\

\rowcolor{rowshade}
\emph{Charlton: Nothing About Us Without Us (1998)} &
Establishes the disability rights principle that affected populations must lead decisions about their own lives \\

\emph{Rose: Governing the Soul (1989)} &
Analyses how psy-disciplines extend institutional power through the governance of subjectivity \\

\rowcolor{rowshade}
\emph{AI Now Institute: Discriminating Systems (2019)} &
Documents gender, race and power in AI systems as structural rather than incidental \\

\bottomrule
\end{tabular}
\end{table*}

\begin{table*}[!htbp]
\centering
\small
\caption{AI and neurodivergence. Representative works on neurodivergent engagement with AI systems and the cognitive costs of normative design.}
\label{tab:neurodivergence}
\setlength{\tabcolsep}{6pt}
\renewcommand{\arraystretch}{1.35}
\begin{tabular}{>{\raggedright\arraybackslash}p{4.8cm} >{\raggedright\arraybackslash}p{9.4cm}}
\arrayrulecolor{headershade}
\toprule
\rowcolor{headershade}
\textcolor{white}{\textbf{Work}} & \textcolor{white}{\textbf{Contribution}} \\
\midrule

\rowcolor{rowshade}
\emph{Cage \& Troxell-Whitman: Understanding the Reasons, Contexts and Costs of Camouflaging (2019)} &
Empirically maps the reasons, contexts and costs of camouflaging for autistic adults \\

\emph{Papadopoulos: Large Language Models for Autistic and Neurodivergent Individuals (2024)} &
Critically examines LLM affordances and risks for the neurodivergent community through personal experimentation and community discussion \\

\rowcolor{rowshade}
\emph{Xue et al.: Characterizing Uses and Prompting Strategies of LLM-Based Chatbots (2025)} &
Interview and diary-study methodology capturing real-life chatbot use and prompting strategies of neurodivergent users \\

\emph{Jamshed et al.: Rethinking Productivity (2025)} &
Rethinks productivity with generative AI from neurodivergent students' perspectives \\

\rowcolor{rowshade}
\emph{Chapman: Neurodiversity and the Social Ecology of Mental Functions (2021)} &
Ecological functional model reframing neurocognitive diversity as relational rather than deficit-based \\

\emph{Park et al.: The Bias Paradox Around Autism in LLMs (2025)} &
Persona-generation study finding ChatGPT reproduces deficit-oriented stereotypes about autistic people while simultaneously emphasising inclusion and representation \\

\bottomrule
\end{tabular}
\end{table*}

\begin{table*}[!htbp]
\centering
\small
\caption{Design. Representative works on inclusive design, classification systems and the ethical architecture of technology.}
\label{tab:design}
\setlength{\tabcolsep}{6pt}
\renewcommand{\arraystretch}{1.35}
\begin{tabular}{>{\raggedright\arraybackslash}p{4.8cm} >{\raggedright\arraybackslash}p{9.4cm}}
\arrayrulecolor{headershade}
\toprule
\rowcolor{headershade}
\textcolor{white}{\textbf{Work}} & \textcolor{white}{\textbf{Contribution}} \\
\midrule

\rowcolor{rowshade}
\emph{Xu et al.: Cognitive Agency Surrender (2026)} &
Theorises scaffolded cognitive friction as defence against epistemic erosion from zero-friction AI design \\

\emph{Hamraie \& Fritsch: Crip Technoscience (2019)} &
Develops disability-led design as a framework for equitable technology development \\

\rowcolor{rowshade}
\emph{Bennett \& Keyes: What Is the Point of Fairness? (2020)} &
Examines what happens when disability and AI fairness intersect \\

\emph{Chandra et al.: TherapyProbe (2026)} &
Develops adversarial relational-safety probes for mental health chatbots; designs for safety without foreclosing engagement \\

\rowcolor{rowshade}
\emph{Mishra et al.: Understanding AI Guardrails (2025)} &
Develops concepts and methods for understanding AI guardrails as design choices with costs \\

\bottomrule
\end{tabular}
\end{table*}

\fi

\end{document}